\documentclass[twocolumn,showpacs,preprintnumbers,superscriptaddress,floatfix,nofootinbib]{revtex4}
\usepackage{graphicx,subfigure,wrapfig,color}
\usepackage{amsmath,dcolumn,amsmath,amssymb}
\usepackage{bm,longtable}
\begin{document}


\title{EIGHT-CLUSTER STRUCTURE OF CHLOROPLAST GENOMES DIFFERS FROM SIMILAR ONE OBSERVED FOR BACTERIA}

\author{Michael G.Sadovsky}
\affiliation{Institute of computational modelling of SD of RAS;\\ 660036 Russia, Krasnoyarsk,
Akademgorodok.} \email{msad@icm.krasn.ru}
\affiliation{Siberian Federal university;\\ 660041 Russia, Krasnoyarsk,
Svobodny prosp.79.}

\author{Maria Yu.\,Senashova}
\affiliation{Institute of computational modelling of SD of RAS;\\ 660036 Russia, Krasnoyarsk,
Akademgorodok.} \email{msen@icm.krasn.ru}
\author{Andrew V.\,Malyshev}
\affiliation{Institute of computational modelling of SD of RAS;\\ 660036 Russia, Krasnoyarsk,
Akademgorodok.} \email{amal@icm.krasn.ru}
\begin{abstract}
Previously, a seven-cluster pattern claiming to be a universal one in bacterial genomes has been reported. Keeping in mind the most popular theory of chloroplast origin, we checked whether a similar pattern is observed in chloroplast genomes. Surprisingly, eight cluster structure has been found, for chloroplasts. The pattern observed for chloroplasts differs rather significantly, from bacterial one, and from that latter observed for cyanobacteria. The structure is provided by clustering of the fragments of equal length isolated within a genome so that each fragment is converted in triplet frequency dictionary with non-overlapping triplets with no gaps in frame tiling. The points in 63-dimensional space were clustered due to elastic map technique. The eight cluster found in chloroplasts comprises the fragments of a genome bearing tRNA genes and exhibiting excessively high $\mathsf{GC}$-content, in comparison to the entire genome.
\end{abstract}


\maketitle

\section{\label{introd}Introduction}
Detailed study of a structure of nucleotide sequences is a key issue in up-to-date molecular biology and bioinformatics. Such studies are carried out in two (interconnecting) paradigms: the former is structure-function relationship, and the latter is evolutionary one. A retrieval of the interrelation between structure and function of various biological macromoleculae is a core issue of up-to-date molecular and system biology. Currently, a huge number of publications appears annually on this subject; yet, the problem is still far from any completion. Moreover, some new structures are reported nowadays \cite{molevol94,plos1-1}.

Evolutionary value of such studies is rather apparent: comparing various structures found in DNA sequences of various organisms, one expects to retrieve the evolution process details ranging from races and species to global ecological systems. Meanwhile, such studies face a number of problems in selection and quality of biological material to be considered. Skipping off possible errors in sequencing and/or annotation of genetic entities, one faces a great complexity of genomes, or even separate chromosomes. Here one has to study a three-sided entity: structure, function, and phylogeny. Quite often all three issues are so tightly interweaved that one fails to distinguish the effects and contributions of each issue separately.

Prokaryotic organisms seem to be rather suitable for this type of researches: bacterial genome is considerably short and always consists of a single chromosome. An ambiguity in bacterial taxonomy looks like a pay-off for the genome simplicity of these organisms; the problem in taxonomy grows up, as higher taxa are considered \cite{chunrainy,bactsex}. In such capacity, organella genomes seem to be very suitable object for the studies of the type mentioned above: keeping oneself within the organella of the same type (say, chloroplasts), one avoids any problems related to a diversity of functional charge of a genome.

Here we explore the relation between structure and taxonomy of the bearers of chloroplast genomes. A number of papers aims to study evolutionary processes on the basis of genome sequences structures peculiarities retrieval \cite{review12,phylo34} or a comparative study of some peculiar fragments of genomes \cite{curgen96,poljaki,molevol94,plos1-1,amjb2006,amjb2005} of chloroplasts.

Sounding diversity of structures that could be found in DNA sequences is another problem. Surely, the problem hardness depends on the notion of a structure in DNA molecule. Hereafter a structure is stipulated to be a pattern in mutual interlocation of nucleotides manifesting in statistical properties of formally identified short fragments of a genome, i.\,e. the ensemble of strings of the given length~$q$. Further we shall concentrate on the ensembles of strings of the length~$q=3$ (triplets). Henceforth, the list of triplets observed within a genome or its part accompanied with the frequencies of these former is the structure under consideration; see details below.

Indeed, we shall concentrate on the study of mutual location of the points of 63-dimensional space of triplet frequencies, where each point corresponds to a fragment identified within a genome, due to some regular procedure. The matter of interest is a cluster structuredness (if any) of those fragments of genomes converted into frequency dictionaries of triplets, in 63-dimensional metric space. Such approach has been originally explored by Alexander Gorban and co-authors \cite{gorban1,gorban2}, for bacterial genomes. They have found seven-cluster patterns in the fragments distribution, where the specific type of the pattern is strongly ruled by $\mathsf{GC}$-content of a genome.

The most popular theory of chloroplast origin, that is bacterial symbiogenesis theory \cite{merezh,merezh1,cavalier-smith,chlororigin,ravenallen,curopin2014}, stipulates a relation between some bacteria, and chloroplasts; that is the motivation standing behind our study: whether this relation manifests in a similarity of the patterns observed for bacteria \cite{gorban1,gorban2}, and those observed for chloroplasts. Briefly speaking, the answer is negative. Moreover, chloroplast genomes exhibit rather specific patterns drastically differing them from bacterial genomes, and further we shall present the result demonstrating the difference, and discuss this point.

\section{Materials and methods}
Chloroplast genomes were retrieved from EMBL--bank; the list of entities comprises 178 entries. Table~\ref{genomy} enlists the genomes under consideration. Evidently, genomes differ in the quality of sequencing: they may have extra symbols other than $\aleph=\{\mathsf{A},\mathsf{C},\mathsf{G},\mathsf{T}\}$. Wherever it happened, we eliminated such extra symbols from a sequence, and concatenated it into the coherent one.

\subsection{Frequency dictionaries and genome fragmentation}
We stipulate a genome to be a coherent sequence from four-letter alphabet $\aleph=\{\mathsf{A},\mathsf{C},\mathsf{G},\mathsf{T}\}$; the number $N$ of nucleotides is the length of the sequence. Let then fix the length~$q$ of a window, and the length~$t$ of a step. Moving then the window upright (for certainty) alongside the sequence with the step~$t$ and counting the number $n_{\omega}$ of strings $\omega$ of the given length~$q$ identified by the sequential locations of the window, one gets the finite dictionary~$\mathsf{F}_{(q,t)}$. Changing the numbers $n_{\omega}$ for frequencies
\begin{equation}\label{frek}
f_{\omega} = \dfrac{n_{\omega}}{M}\qquad \textrm{with} \qquad M = \sum_{\omega} n_{\omega}\,,
\end{equation}
one gets the frequency dictionary~$W_{(q,t)}$; $M=N$ for $W_{(q,1)}$. Actually, such definition of~$W_{(q,t)}$ requires to connect the sequence under consideration into a ring (see details in \cite{gorban1,gorban2,nasha2}).

Everywhere further we shall constrain with the dictionaries~$W_{(3,3)}$, only. It enlists the triplets counted with no overlapping, while with no gaps between the frames. The choice of $q=3$ and $t=3$ is motivated by apparent biological issues: triplets yield the strongest signal, in DNA sequences, and the step~$t=3$ reveals it, that is coding positions in DNA sequence.

The frequency dictionary~$W_{(3,3)}$ exists in three different (in general) versions differing in reading frame position; that latter is called \textsl{phase of a fragment} below. Indeed, for the sequence
\begin{widetext}
\begin{equation}\label{primer_seq}
\mathsf{CTTGTGCGATCAATTGTTGATCCAGTTTTATGATTGCACCGCAGAAAGTG}
\end{equation}
\end{widetext}
the dictionaries~$W_{(3,3)}$: $W^{(0)}_{(3,3)}$, $W^{(1)}_{(3,3)}$ and $W^{(2)}_{(3,3)}$ are shown in Table~\ref{3slov}. Here the number of copies, not the frequencies are shown.
\begin{table}[!ht]
\centering
\caption{\label{3slov}
{The dictionaries $W^{(0)}_{(3,3)}$, $W^{(1)}_{(3,3)}$ and $W^{(2)}_{(3,3)}$ for the sequence~\eqref{primer_seq}.}}
\begin{tabular}{|l|l|l|l|l|l|l|l|l|l|l|l|}
\hline
\multicolumn{6}{|c|}{$W^{(0)}_{(3,3)}$} & \multicolumn{6}{c|}{$W^{(1)}_{(3,3)}$}\\\hline
$\mathsf{AAG}$&1&$\mathsf{CGA}$&1&$\mathsf{GTT}$&2&$\mathsf{ACC}$&1&$\mathsf{GAA}$&1&$\mathsf{TTG}$&3\\
$\mathsf{AGA}$&1&$\mathsf{CGC}$&1&$\mathsf{TCA}$&1&$\mathsf{AGT}$&1&$\mathsf{GAT}$&2&$\mathsf{TTT}$&1\\
$\mathsf{ATT}$&1&$\mathsf{CTT}$&1&$\mathsf{TGA}$&1&$\mathsf{ATC}$&1&$\mathsf{GCA}$&1&&\\\hline
$\mathsf{CAC}$&1&$\mathsf{GAT}$&1&$\mathsf{TTA}$&1&$\mathsf{CAA}$&1&$\mathsf{TAT}$&1&&\\\hline
$\mathsf{CCA}$&1&$\mathsf{GTG}$&1&$\mathsf{TTG}$&1&$\mathsf{CAG}$&1&$\mathsf{TGC}$&2&&\\\hline
\multicolumn{6}{|c|}{$W^{(2)}_{(3,3)}$}\\\cline{1-6}
$\mathsf{AAA}$&1&$\mathsf{ATT}$&1&$\mathsf{GTG}$&1\\\cline{1-6}
$\mathsf{AAT}$&1&$\mathsf{CAG}$&1&$\mathsf{TCC}$&1\\\cline{1-6}
$\mathsf{AGT}$&1&$\mathsf{CCG}$&1&$\mathsf{TGA}$&1\\\cline{1-6}
$\mathsf{ATC}$&1&$\mathsf{GCA}$&1&$\mathsf{TGT}$&2\\\cline{1-6}
$\mathsf{ATG}$&1&$\mathsf{GCG}$&1&$\mathsf{TTT}$&1\\\cline{1-6}
\end{tabular}
\end{table}
Everywhere below, we shall develop the frequency dictionaries $W^{(0)}_{(3,3)}$ for each fragment of a sequence. Reciprocally, the \textsl{phase} of the fragment has been determined, instead of the implementation of two other dictionaries; see details below.

To figure out the inner structuredness of a chloroplast genome, we cut it into a set of (overlapping) fragments. To do that, the length of a fragment~$L$ and the move step~$R$ alongside a genomes have been fixed; we used the figures $L=603$ and $R=11$, in our studies. The motivation for the choice of such figures is following: we need to choose the length~$L$ of a fragment to be odd and divisible by~3, while the step~$R$ must be not divisible by~3. Next, the length of a fragment is chosen rather close to a gene length. The step length~$R$ determines the number of points taken into consideration, e.\,g. for $K$-means clustering; the chosen step figure yields $\sim 10^4$ fragments (later converted into the points in a metric space). Obviously, both~$L$ and~$R$ could be set \textit{de novo}, if necessary.

Any frequency dictionary $W_{(3,\cdot )}$ maps a sequence into a point in 63-dimensional space. Indeed, the total number of triplets is equal to 64; meanwhile, the linear constraint
\begin{equation}\label{ediniza}
\sum_{\omega = \mathsf{AAA}}^{\mathsf{TTT}} f_{\omega} = 1
\end{equation}
makes remain only 63 ones independent; the frequency of the last one is unambiguously determined from~\eqref{ediniza}. Formally, any triplet may be eliminated; practically, we excluded the triplet exhibiting the least standard deviation figure determined over the entire ensemble of the fragments. Table~\ref{genomy} shows these triplets, at the column labeled $\omega_{\min}$.

Apparently, there might be other ways to determine the excluded triplet. For example, it is useful to exclude the variable with maximum value, for some situations; here we followed the described above way, since the least standard deviation of a triplet frequencies observed over a dataset means the least distinguishability of the objects comprising a dataset, over this variable. Thus, the dimensionality of the space to cluster the frequency dictionaries of triplets becomes equal to~63.

\subsubsection{The phase of a fragment of sequence}
Previously, three types of frequency dictionaries $W^{(0)}_{(3,3)}$, $W^{(1)}_{(3,3)}$ and $W^{(2)}_{(3,3)}$ were shown (see Table~\ref{3slov}). Meanwhile, we developed only one frequency dictionary; that was $W^{(0)}_{(3,3)}$ dictionary. The fragment was then labeled using one of four labels: \textsl{phase}~1, \textsl{phase}~2, \textsl{phase}~3 and \textsl{junk}. The label was determined by the location of a fragment within a sequence; to do that, we used the annotation of each genome under consideration.\medskip

A fragment was labeled as
\begin{description}
\item[\textsl{junk,}] if it contains at least a half of a non-coding region of a genome within itself;\label{junkdef}
\item[\textsl{phase}~0,] if the center of a fragment falls into a coding region of a genome, and the reminder of the division of the distance between the central nucleotide of a fragment, and the starting nucleotide of a coding region is equal to~0;
\item[\textsl{phase}~1,] if the center of a fragment falls into a coding region of a genome, and the reminder of the division of the distance between the central nucleotide of a fragment, and the starting nucleotide of a coding region is equal to~1;
\item[\textsl{phase}~2,] if the center of a fragment falls into a coding region of a genome, and the reminder of the division of the distance between the central nucleotide of a fragment, and the starting nucleotide of a coding region is equal to~2.
\end{description}
For genes (or coding regions) located in the ladder strand, the above mentioned procedure still holds true, but the distance to the central nucleotide of a fragment is determined not from the start position (formally indicated in a file), but from the end of that latter.

\subsection{Clustering of frequency dictionaries}
As soon, as the fragments are converted into the frequency dictionaries $W^{(0)}_{(3,3)}$, then each dictionary was labeled with the number of the nucleotide occupying the central position at the corresponding fragment. Also, each fragment was labeled with its \textsl{phase}. To make the space of frequency dictionaries metric, one must implement a metrics; there is a number of options here (see \cite{c3,obzor1,n3_3,n4,gorbanzyn2,gorbanzyn,gorbanzyn1,gorbanzyn3} for details). Meanwhile, we use Euclidean metrics:
\begin{equation}\label{evklid}
\rho\left(W^{[1]}_{(3,3)}, W^{[2]}_{(3,3)}\right) = \sqrt{\sum_{\omega=\mathsf{AAA}}^{\mathsf{TTT}} \left( f^{[1]}_{\omega} - f^{[2]}_{\omega} \right)^2}\,.
\end{equation}
Here $f^{[j]}_{\omega}$ is the frequency of a triplet $\omega$ observed in the $j^{\textrm{th}}$~frequency dictionary; this index has nothing to do with the frame shift described above.

We studied the distribution of those fragments, in 63-dimensional space using \textsl{VidaExpert} (\textit{http://bioinfo-out.curie.fr/projects/vidaexpert/}) software. No special technique for clustering has been used: we identified the clusters \textit{as is}, through visualization. Nonetheless, all the clusters identified through visualization were also identified with $K$-means; thus, those clusters could be verified objectively.

In addition, $\mathsf{GC}$-content has been determined, both for each fragment, and the genome entirely (see Table~\ref{genomy}, $\mathsf{GC}$ labeled column).


\section{Results}
First, let's consider the list of chloroplast genomes used in the study, in more detail. The list is quite homogeneous, in terms of the length of sequences; thus, we may not expect any effect resulted from a length difference. Next point is the eliminated triplet choice; Table~\ref{genomy} shows those triplets at the sixth column. Actually, there are only four triplets eliminated in various genomes: $\mathsf{CGC}$ (58 entries), $\mathsf{GCG}$ (113 entries), $\mathsf{GAC}$ (1 entry) and $\mathsf{TAA}$ (also 1 entry).

The triplets $\mathsf{CGC}$ and $\mathsf{GCG}$ are of great interest: they both are palindromes (read equally in opposite directions), and besides they together comprise the couple of so called \textsl{complementary palindrome}. That latter consists of two string (triplets, in our case) that are read equally in opposite directions, with respect to Chargaff's parity rule: $\mathsf{CGC} \Leftrightarrow \mathsf{GCG}$. Such symmetry is rather important both in analysis, and in biological issues standing behind it; more detailed discussion see below.\vfill

\begin{longtable*}{|p{4.4cm}|r|c|c|c|c|c|c|}
\caption{\label{genomy} List of genomes studied; $N$ is the length of genome, $\mathsf{GC}$ is $\mathsf{GC}$-content, $J$ is junk percentage, and $\omega_{\min}$ is the triplet with minimal standard deviation}\\
\hline \multicolumn{1}{|c|}{Genomes}& AC number & \multicolumn{1}{c|}{$N$} & $\mathsf{GC}$ & $J$ & $\omega_{\min}$ & $\Updownarrow$\\
\hline
\endfirsthead
\multicolumn{7}{r}%
{{\tablename\ \thetable{} -- continued}} \\
\hline \multicolumn{1}{|c|}{Genomes}& AC number & \multicolumn{1}{c|}{$N$} & $\mathsf{GC}$ & $J$ & $\omega_{\min}$ & $\Updownarrow$\\
\hline
\endhead
\hline \multicolumn{7}{|r|}{{continued on the next page}} \\ \hline
\endfoot
\endlastfoot
Allium cepa (onion) & KF728080 & 153538 & 0.37 & 41.84 & CGC & $\mathsf{U}$\\\hline
Aneura mirabilis & EU043314 & 108007 & 0.41 & 56.11 & GCG & $\mathsf{U}$\\\hline
Angiopteris evecta & DQ821119 & 153901 & 0.35 & 54.15 & GCG & $\mathsf{U}$\\\hline
Anthoceros angustus & AB086179 & 161162 & 0.33 & 48.48 & GCG & $\mathsf{D}$\\\hline
Apopellia endiviifolia & JX827163 & 120544 & 0.36 & 38.62 & CGC & $\mathsf{U}$\\\hline
Arabidopsis thaliana & AP000423 & 154478 & 0.36 & 48.76 & CGC & $\mathsf{U}$\\\hline
Brachypodium distachyon & EU325680 & 135197 & 0.39 & 59.56 & GCG & $\mathsf{D}$\\\hline
Cycas revoluta & JN867588 & 162489 & 0.39 & 45.93 & CGC & $\mathsf{D}$\\\hline
Equisetum arvense & GU191334 & 133309 & 0.33 & 45.07 & GCG & $\mathsf{U}$\\\hline
Fagopyrum esculentum & EU254477 & 159596 & 0.38 & 47.91 & GCG & $\mathsf{U}$\\\hline
Fargesia nitida & JX513416 & 139535 & 0.39 & 57.45 & GCG & $\mathsf{D}$\\\hline
Fragaria chiloensis & JN884816 & 155603 & 0.37 & 48.83 & GCG & $\mathsf{U}$\\\hline
Fritillaria hupehensis & KF712486 & 152145 & 0.37 & 52.61 & GCG & $\mathsf{U}$\\\hline
Gaoligongshania megalothyrsa & JX513419 & 140064 & 0.39 & 57.58 & GCG & $\mathsf{U}$\\\hline
Genlisea margaretae & HG530134 & 141252 & 0.38 & 58.61 & GCG & $\mathsf{U}$\\\hline
Ginkgo biloba & AB684440 & 156945 & 0.40 & 53.00 & GCG & $\mathsf{U}$\\\hline
Glycine max & DQ317523 & 152218 & 0.35 & 48.49 & CGC & $\mathsf{U}$\\\hline
Glycyrrhiza glabra & KF201590 & 127942 & 0.34 & 47.78 & CGC & $\mathsf{D}$\\\hline
Gnetum montanum & KC427271 & 115019 & 0.38 & 46.70 & CGC & $\mathsf{U}$\\\hline
Goodyera fumata & KJ501999 & 155643 & 0.37 & 48.37 & CGC & $\mathsf{U}$\\\hline
Gossypium anomalum & JF317351 & 159505 & 0.37 & 50.02 & GCG & $\mathsf{U}$\\\hline
Guizotia abyssinica & EU549769 & 151762 & 0.38 & 48.47 & CGC & $\mathsf{D}$\\\hline
Habenaria pantlingiana & KJ524104 & 153951 & 0.37 & 48.25 & GCG & $\mathsf{U}$\\\hline
Helianthus annuus & DQ383815 & 151104 & 0.38 & 48.55 & GCG & $\mathsf{U}$\\\hline
Hibiscus syriacus & KP688069 & 161019 & 0.37 & 60.43 & GCG & $\mathsf{U}$\\\hline
Hordeum vulgare & KC912688 & 114434 & 0.37 & 50.92 & CGC & $\mathsf{D}$\\\hline
Huperzia lucidula & AY660566 & 154373 & 0.36 & 52.05 & CGC & $\mathsf{U}$\\\hline
Hyoscyamus niger & KF248009 & 155720 & 0.38 & 48.91 & CGC & $\mathsf{U}$\\\hline
Hypseocharis bilobata & KF240616 & 165002 & 0.39 & 49.61 & CGC & $\mathsf{U}$\\\hline
Illicium oligandrum & EF380354 & 148553 & 0.39 & 52.06 & GCG & $\mathsf{U}$\\\hline
Indosasa sinica & JX513422 & 139660 & 0.39 & 57.70 & CGC & $\mathsf{U}$\\\hline
Iochroma nitidum & KP294386 & 156574 & 0.38 & 50.61 & CGC & $\mathsf{U}$\\\hline
Ipomoea batatas & KP212149 & 161303 & 0.38 & 53.18 & GCG & $\mathsf{U}$\\\hline
Jacobaea vulgaris & HQ234669 & 150688 & 0.37 & 49.17 & GCG & $\mathsf{U}$\\\hline
Jasminum nudiflorum & DQ673255 & 165121 & 0.38 & 49.87 & GCG & $\mathsf{U}$\\\hline
Jatropha curcas & FJ695500 & 163856 & 0.35 & 51.33 & GCG & $\mathsf{U}$\\\hline
Juniperus bermudiana & KF866297 & 127631 & 0.35 & 40.85 & CGC & $\mathsf{U}$\\\hline
Juniperus monosperma & KF866298 & 127744 & 0.35 & 40.75 & CGC & $\mathsf{U}$\\\hline
Juniperus virginiana & KF866300 & 127770 & 0.35 & 40.92 & CGC & $\mathsf{U}$\\\hline
Kalopanax septemlobus & KC456167 & 156413 & 0.38 & 50.68 & GCG & $\mathsf{U}$\\\hline
Keteleeria davidiana & AP010820 & 117720 & 0.39 & 45.19 & CGC & $\mathsf{U}$\\\hline
Lactuca sativa & AP007232 & 152765 & 0.38 & 52.30 & GCG & $\mathsf{U}$\\\hline
Larix decidua & AB501189 & 122474 & 0.39 & 49.58 & CGC & $\mathsf{U}$\\\hline
Lathyrus sativus & HM029371 & 121020 & 0.35 & 44.25 & GCG & $\mathsf{U}$\\\hline
Lecomtella madagascariensis & HF543599 & 139073 & 0.39 & 57.13 & GCG & $\mathsf{U}$\\\hline
Lemna minor & DQ400350 & 165955 & 0.36 & 49.42 & GCG & $\mathsf{U}$\\\hline
Lepidium virginicum & AP009374 & 154743 & 0.36 & 48.68 & GCG & $\mathsf{U}$\\\hline
Licania alba & KJ414483 & 162467 & 0.36 & 51.53 & GCG & $\mathsf{U}$\\\hline
Lindenbergia philippensis & HG530133 & 155103 & 0.38 & 48.93 & GCG & $\mathsf{U}$\\\hline
Liquidambar formosana & KC588388 & 160410 & 0.38 & 50.85 & GCG & $\mathsf{U}$\\\hline
Liriodendron tulipifera & DQ899947 & 159886 & 0.39 & 50.29 & GCG & $\mathsf{U}$\\\hline
Lobularia maritima & AP009375 & 152659 & 0.37 & 48.37 & GCG & $\mathsf{U}$\\\hline
Lolium perenne & AM777385 & 135282 & 0.38 & 55.66 & GCG & $\mathsf{U}$\\\hline
Lonicera japonica & KJ170923 & 155078 & 0.39 & 51.72 & CGC & $\mathsf{U}$\\\hline
Lotus japonicus & AP002983 & 150519 & 0.36 & 48.85 & CGC & $\mathsf{U}$\\\hline
Lupinus luteus & KC695666 & 151891 & 0.37 & 48.73 & GCG & $\mathsf{U}$\\\hline
Lygodium japonicum & KC536645 & 157260 & 0.41 & 48.69 & CGC & $\mathsf{U}$\\\hline
Magnolia kwangsiensis & HM775382 & 159667 & 0.39 & 50.92 & GCG & $\mathsf{U}$\\\hline
Manihot esculenta & EU117376 & 161453 & 0.36 & 55.23 & GCG & $\mathsf{U}$\\\hline
Mankyua chejuensis & JF343520 & 146221 & 0.38 & 52.36 & CGC & $\mathsf{U}$\\\hline
Marchantia paleacea & X04465 & 121024 & 0.29 & 39.93 & GCG & $\mathsf{D}$\\\hline
Marsilea crenata & KC536646 & 151628 & 0.42 & 47.23 & GCG & $\mathsf{U}$\\\hline
Masdevallia coccinea & KP205432 & 157423 & 0.37 & 49.71 & GCG & $\mathsf{U}$\\\hline
Megaleranthis saniculifolia & FJ597983; & 159924 & 0.38 & 50.77 & GCG & $\mathsf{U}$\\\hline
Metapanax delavayi & KC456165 & 156343 & 0.38 & 50.65 & GCG & $\mathsf{U}$\\\hline
Millettia pinnata & JN673818 & 152968 & 0.35 & 49.03 & CGC & $\mathsf{U}$\\\hline
Morus indica & DQ226511 & 158484 & 0.36 & 50.05 & GCG & $\mathsf{U}$\\\hline
Myriopteris lindheimeri & HM778032 & 155770 & 0.43 & 48.17 & GCG & $\mathsf{U}$\\\hline
Nageia nagi & AB830885 & 133722 & 0.37 & 44.15 & GCG & $\mathsf{D}$\\\hline
Najas flexilis & JX978472 & 156362 & 0.38 & 58.16 & GCG & $\mathsf{D}$\\\hline
Nasturtium officinale & AP009376 & 155105 & 0.36 & 49.00 & CGC & $\mathsf{U}$\\\hline
Nelumbo lutea & FJ754269 & 163206 & 0.38 & 51.25 & GCG & $\mathsf{U}$\\\hline
Neyraudia reynaudiana & KF356392 & 135367 & 0.38 & 55.66 & CGC & $\mathsf{U}$\\\hline
Nicotiana sylvestris & AB237912 & 155941 & 0.38 & 46.15 & CGC & $\mathsf{U}$\\\hline
Nuphar advena & DQ354691 & 160866 & 0.39 & 50.64 & GCG & $\mathsf{U}$\\\hline
Nymphaea alba & AJ627251 & 159930 & 0.39 & 49.50 & GCG & $\mathsf{U}$\\\hline
Oenothera argillicola & EU262887 & 165061 & 0.39 & 50.34 & CGC & $\mathsf{U}$\\\hline
Olea europaea & GU228899 & 155888 & 0.38 & 49.06 & GCG & $\mathsf{U}$\\\hline
Oligostachyum shiuyingianum & JX513423 & 139647 & 0.39 & 57.64 & CGC & $\mathsf{U}$\\\hline
Olimarabidopsis pumila & AP009368 & 154737 & 0.36 & 49.16 & GCG & $\mathsf{U}$\\\hline
Ophioglossum californicum & KC117178 & 138270 & 0.42 & 47.57 & CGC & $\mathsf{U}$\\\hline
Orobanche gracilis & HG803179 & 65533 & 0.35 & 73.30 & GCG & $\mathsf{U}$\\\hline
Orthotrichum rogeri & KP119739 & 123363 & 0.28 & 43.01 & CGC & $\mathsf{D}$\\\hline
Oryza nivara & AP006728 & 134494 & 0.39 & 50.23 & CGC & $\mathsf{D}$\\\hline
Oryza sativa & JN861109 & 134448 & 0.39 & 54.61 & CGC & $\mathsf{D}$\\\hline
Oryza sativa & JN861110 & 134459 & 0.39 & 55.21 & CGC & $\mathsf{D}$\\\hline
Oryza sativa Indica & AY522329 & 134496 & 0.39 & 63.88 & CGC & $\mathsf{D}$\\\hline
Oryza sativa Japonica & AY522330 & 134551 & 0.39 & 66.96 & GCG & $\mathsf{D}$\\\hline
Oryza sativa Japonica & GU592207 & 134551 & 0.39 & 64.36 & GCG & $\mathsf{D}$\\\hline
Oryza sativa Japonica & X15901 & 134525 & 0.39 & 42.96 & CGC & $\mathsf{D}$\\\hline
Pachycladon cheesemanii & JQ806762 & 154498 & 0.36 & 49.20 & GCG & $\mathsf{U}$\\\hline
Paeonia obovata & KJ206533 & 152696 & 0.38 & 49.68 & GCG & $\mathsf{U}$\\\hline
Panax ginseng & AY582139 & 156313 & 0.38 & 49.32 & GCG & $\mathsf{U}$\\\hline
Panicum virgatum & HQ731441 & 139677 & 0.39 & 57.05 & GCG & $\mathsf{U}$\\\hline
Paphiopedilum armeniacum & KJ566307 & 162682 & 0.35 & 55.30 & CGC & $\mathsf{D}$\\\hline
Parinari campestris & KJ414486 & 162637 & 0.36 & 51.58 & GCG & $\mathsf{U}$\\\hline
Parthenium argentatum & GU120098 & 152803 & 0.38 & 77.21 & CGC & $\mathsf{D}$\\\hline
Pelargonium $\times$  hortorum & DQ897681 & 217942 & 0.40 & 48.14 & GCG & $\mathsf{D}$\\\hline
Pentactina rupicola & JQ041763 & 156612 & 0.37 & 49.40 & GCG & $\mathsf{U}$\\\hline
Penthorum chinense & JX436155 & 156686 & 0.37 & 51.26 & GCG & $\mathsf{U}$\\\hline
Phalaenopsis equestris & JF719062 & 148959 & 0.37 & 56.71 & GCG & $\mathsf{U}$\\\hline
Pharus lappulaceus & KC311467 & 141928 & 0.38 & 58.51 & CGC & $\mathsf{U}$\\\hline
Phoenix dactylifera & GU811709 & 158462 & 0.37 & 50.00 & GCG & $\mathsf{U}$\\\hline
Phragmites australis & KF730315 & 137561 & 0.39 & 56.39 & CGC & $\mathsf{U}$\\\hline
Phyllostachys edulis & HQ337796 & 139679 & 0.39 & 57.14 & GCG & $\mathsf{D}$\\\hline
Phyllostachys propinqua & JN415113 & 139704 & 0.39 & 57.54 & GCG & $\mathsf{D}$\\\hline
Physcomitrella patens & AP005672 & 122890 & 0.29 & 42.08 & CGC & $\mathsf{U}$\\\hline
Picea abies & HF937082 & 124084 & 0.39 & 51.11 & CGC & $\mathsf{U}$\\\hline
Pinguicula ehlersiae & HG803178 & 147140 & 0.38 & 56.14 & GCG & $\mathsf{U}$\\\hline
Pinus contorta & EU998740 & 115267 & 0.38 & 49.93 & CGC & $\mathsf{U}$\\\hline
Pinus taeda (loblolly pine) & KC427273 & 121530 & 0.39 & 44.26 & CGC & $\mathsf{D}$\\\hline
Piper cenocladum & DQ887677 & 160624 & 0.38 & 52.93 & GCG & $\mathsf{U}$\\\hline
Pisum sativum & HM029370 & 122169 & 0.35 & 46.08 & GCG & $\mathsf{U}$\\\hline
Pleioblastus maculatus & JX513424 & 139720 & 0.39 & 57.73 & GCG & $\mathsf{D}$\\\hline
Podocarpus lambertii & KJ010812 & 133734 & 0.37 & 44.25 & GCG & $\mathsf{D}$\\\hline
Populus alba & AP008956 & 156505 & 0.37 & 49.21 & GCG & $\mathsf{U}$\\\hline
Premna microphylla & KM981744 & 155293 & 0.38 & 48.79 & GCG & $\mathsf{U}$\\\hline
Primula poissonii & KF753634 & 151663 & 0.37 & 47.61 & CGC & $\mathsf{U}$\\\hline
Prunus kansuensis & KF990036 & 157736 & 0.37 & 49.51 & GCG & $\mathsf{U}$\\\hline
Pseudotsuga sinensis & AB601120 & 122513 & 0.39 & 50.16 & CGC & $\mathsf{D}$\\\hline
Psilotum nudum & AP004638 & 138829 & 0.36 & 44.90 & GCG & $\mathsf{U}$\\\hline
Pteridium aquilinum & HM535629 & 152362 & 0.42 & 46.95 & CGC & $\mathsf{U}$\\\hline
Ptilidium pulcherrimum & HM222519 & 119003 & 0.33 & 50.36 & CGC & $\mathsf{D}$\\\hline
Puelia olyriformis & KC534841 & 140318 & 0.39 & 57.50 & CGC & $\mathsf{U}$\\\hline
Quercus aliena & KP301144 & 160921 & 0.37 & 51.40 & GCG & $\mathsf{U}$\\\hline
Ranunculus macranthus & DQ359689 & 155129 & 0.38 & 49.02 & GCG & $\mathsf{U}$\\\hline
Raphanus sativus & KJ716483 & 153368 & 0.36 & 47.90 & GCG & $\mathsf{U}$\\\hline
Retrophyllum piresii & KJ617081 & 133291 & 0.37 & 43.87 & CGC & $\mathsf{D}$\\\hline
Rhazya stricta & KJ123753 & 154736 & 0.38 & 49.56 & GCG & $\mathsf{U}$\\\hline
Rosa odorata var. gigantea & KF753637 & 156634 & 0.37 & 49.51 & GCG & $\mathsf{U}$\\\hline
Saccharum hybrid & AP006714 & 141182 & 0.38 & 52.13 & GCG & $\mathsf{U}$\\\hline
Salix interior & KJ742926 & 156488 & 0.37 & 50.78 & GCG & $\mathsf{U}$\\\hline
Salvia miltiorrhiza & HF586694 & 151332 & 0.38 & 46.85 & GCG & $\mathsf{U}$\\\hline
Sanionia uncinata & KM111545 & 124374 & 0.29 & 43.25 & CGC & $\mathsf{D}$\\\hline
Sarocalamus faberi & JX513414 & 139629 & 0.39 & 57.13 & GCG & $\mathsf{D}$\\\hline
Schefflera delavayi & KC456166 & 156340 & 0.38 & 52.95 & GCG & $\mathsf{U}$\\\hline
Schwalbea americana & HG738866 & 160908 & 0.38 & 51.23 & GCG & $\mathsf{U}$\\\hline
Sedum sarmentosum & JX427551 & 150447 & 0.38 & 47.96 & CGC & $\mathsf{U}$\\\hline
Selaginella moellendorffii & HM173080 & 143775 & 0.51 & 50.84 & GAC & $\mathsf{D}$\\\hline
Selaginella uncinata & AB197035 & 144170 & 0.55 & 51.75 & TAA & $\mathsf{D}$\\\hline
Sesamum indicum & JN637766 & 153324 & 0.38 & 48.61 & GCG & $\mathsf{U}$\\\hline
Setaria italica & KJ001642 & 138833 & 0.39 & 56.32 & GCG & $\mathsf{U}$\\\hline
Silene chalcedonica & KF527886 & 148081 & 0.36 & 48.80 & GCG & $\mathsf{U}$\\\hline
Sorghum bicolor & EF115542 & 140754 & 0.38 & 57.71 & GCG & $\mathsf{U}$\\\hline
Spirodela polyrhiza & JN160603 & 168788 & 0.36 & 50.21 & GCG & $\mathsf{U}$\\\hline
Stangeria eriopus & JX416858 & 163548 & 0.40 & 53.62 & CGC & $\mathsf{U}$\\\hline
Stockwellia quadrifida & KC180807 & 159561 & 0.37 & 50.36 & GCG & $\mathsf{U}$\\\hline
Syntrichia ruralis & FJ546412 & 122501 & 0.28 & 45.76 & CGC & $\mathsf{D}$\\\hline
Taxus mairei & KJ123824 & 129521 & 0.35 & 43.14 & CGC & $\mathsf{U}$\\\hline
Tetracentron sinense & KC608752 & 164467 & 0.38 & 49.29 & GCG & $\mathsf{U}$\\\hline
Thamnocalamus spathiflorus & JX513425 & 139778 & 0.39 & 57.47 & CGC & $\mathsf{D}$\\\hline
Trachelium caeruleum & EU090187 & 162314 & 0.38 & 55.90 & CGC & $\mathsf{U}$\\\hline
Trifolium subterraneum & EU849487 & 144763 & 0.34 & 55.12 & GCG & $\mathsf{U}$\\\hline
Trigonobalanus doichangensis & KF990556 & 159938 & 0.37 & 54.96 & GCG & $\mathsf{U}$\\\hline
Triticum aestivum & KC912694 & 114984 & 0.37 & 51.40 & CGC & $\mathsf{D}$\\\hline
Triticum aestivum & KJ592713 & 133873 & 0.38 & 55.70 & GCG & $\mathsf{D}$\\\hline
Triticum aestivum & AB042240 & 134545 & 0.38 & 55.23 & GCG & $\mathsf{U}$\\\hline
Trochodendron aralioides & KC608753 & 165936 & 0.38 & 50.46 & GCG & $\mathsf{U}$\\\hline
Typha latifolia & GU195652 & 161572 & 0.37 & 51.12 & GCG & $\mathsf{U}$\\\hline
Utricularia gibba & KC997777 & 152044 & 0.38 & 49.75 & GCG & $\mathsf{U}$\\\hline
Vaccinium macrocarpon & JQ248601 & 176037 & 0.37 & 72.42 & GCG & $\mathsf{U}$\\\hline
Veratrum patulum & KF437397 & 153699 & 0.38 & 49.23 & GCG & $\mathsf{U}$\\\hline
Vigna radiata & GQ893027 & 151271 & 0.35 & 48.88 & CGC & $\mathsf{U}$\\\hline
Vitis rotundifolia & KF976463 & 160891 & 0.37 & 51.24 & GCG & $\mathsf{U}$\\\hline
Vitis vinifera & AB856289 & 160927 & 0.37 & 51.20 & GCG & $\mathsf{U}$\\\hline
Vitis vinifera & AB856290 & 160927 & 0.37 & 51.18 & GCG & $\mathsf{U}$\\\hline
Vitis vinifera & AB856291 & 160906 & 0.37 & 51.21 & GCG & $\mathsf{U}$\\\hline
Vitis vinifera & DQ424856 & 160928 & 0.37 & 50.93 & GCG & $\mathsf{U}$\\\hline
Viviania marifolia & KF240615 & 157291 & 0.38 & 59.61 & GCG & $\mathsf{U}$\\\hline
Welwitschia mirabilis & EU342371 & 119726 & 0.37 & 44.02 & GCG & $\mathsf{U}$\\\hline
Wolffia australiana & JN160605 & 168704 & 0.36 & 43.94 & GCG & $\mathsf{U}$\\\hline
Yushania levigata & JX513426 & 139633 & 0.39 & 57.41 & GCG & $\mathsf{D}$\\\hline
Zamia furfuracea & JX416857 & 164953 & 0.40 & 50.63 & GCG & $\mathsf{U}$\\\hline
Zea mays & KF241981 & 140447 & 0.38 & 57.32 & GCG & $\mathsf{D}$\\\hline
Zea mays & X86563 & 140384 & 0.38 & 43.85 & GCG & $\mathsf{D}$\\\hline
Zea mays & KF241980 & 140437 & 0.38 & 57.34 & GCG & $\mathsf{U}$\\\hline
\end{longtable*}


\subsection{Eight cluster structure of chloroplast genomes}
Let now consider the patterns of chloroplast genomes. To do that, we just located the points corresponding to frequency dictionaries of the fragments of a chloroplast genome, in 63-dimensional space. Of course, there is no way to see this distribution immediately. We used \textsl{ViDaExpert} software \cite{vida} to visualize it. The best way to see a pattern provided by distribution of the fragments converted into frequency dictionaries is to see it in the space determined by three main principal components \cite{fukunaga}.

To begin with, we shall expand the labeling system described above. Previously, four labels have been introduced: \textsl{phase}~0, \textsl{phase}~1, \textsl{phase}~2 and \textsl{junk}. Now we add one more phase called \textsl{tail}, and split each \textsl{phase}~$j^{\textrm{th}}$ into two subphases: these are the phases $F_0$, $F_1$, $F_2$, and $B_0$, $B_1$, $B_2$, respectively. The sense of these subphases is clear and apparent: they correspond to forward reading ($F_0$, $F_1$ and $F_2$) and backward reading ($B_0$, $B_1$ and $B_2$) of genes, in leader and ladder strands, respectively. The index coincides to the reminder of the division of the distance between the start position of a coding regions, and the center of a fragment, by~3.

The \textsl{tail} phase looks the most intriguing. First of all, it comprises the fragments falling into a dense series of tRNA (5S\,RNA, 25S\,RNA, etc.) genes. Probably, the points indicated as \textsl{junk} in the \textsl{tail}~phase are the border fragments.

Now consider several genomes shown in principal components in two projection: in ``full face'' and in ``profile''. The former means that the first principal component is normal to the plane of view, and the latter means that the first principal component is in the plane of view.
\begin{figure*}
\centering
\includegraphics[width=0.47\textwidth]{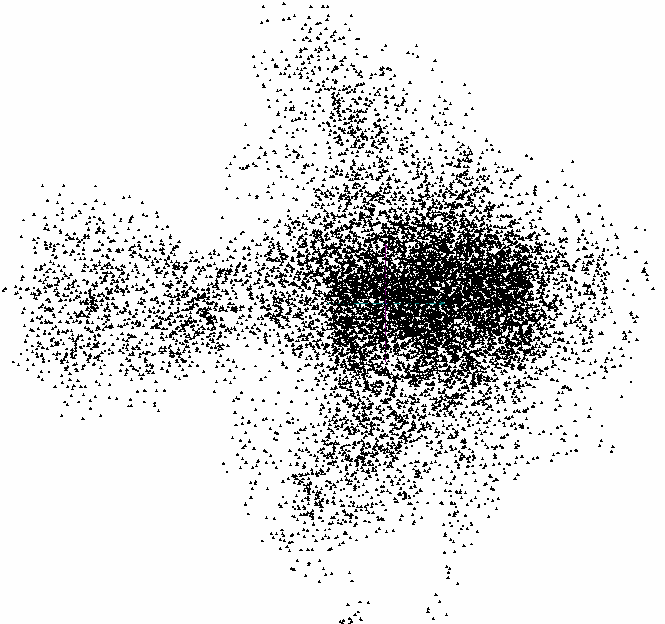}\hfill
\includegraphics[width=0.47\textwidth]{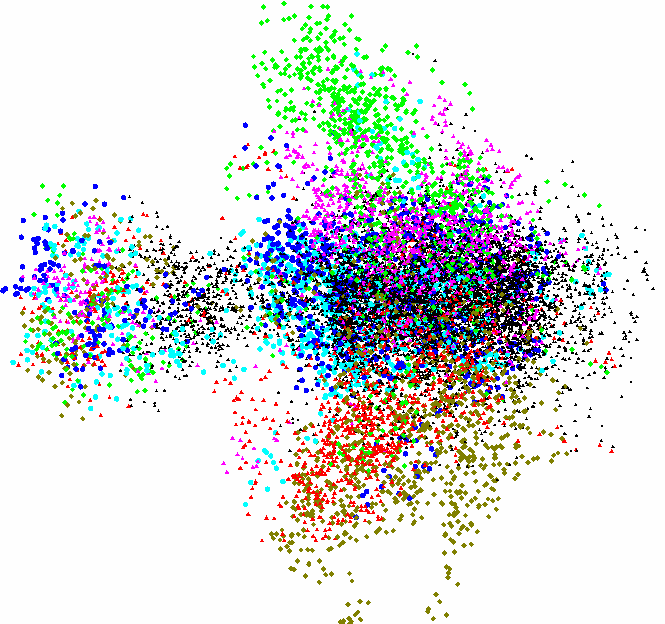}
\caption{{The distribution of 12\,244 fragments of \textit{Lolium perenne} chloroplast genome (AC\,AM777385), ``profile'', in principle coordinates. Left image presents a general overview, and the right one presents the phases $F_0$~-- red triangles, $F_1$~-- bright green diamonds, $F_2$~-- light blue circles, $B_0$~-- rosy triangles, $B_1$~-- sand diamonds, $B_2$~-- dark blue circles.}}
\label{fig1}
\end{figure*}
Fig.~\ref{fig1} presents the ``profile'' view of the fragments distribution of ray grass (\textit{Lolium perenne}, AC\,AM777385 in EMBL--bank) genome, the coding regions. This is a typical ``bullet-like'' pattern of the distribution. In the right in this Figure the same distribution is shown with indication of the fragments labeled in color (see Figure legend). \textsl{Junk} phase is shown in the right figure in black.

Thus, eight clusters are distinctively identified, in this Figure: six ones correspond to six phases (from $F_0$ to $B_2$, respectively), the seventh cluster comprises the \textsl{junk} labeled fragments, and the eighth cluster (that is the \textsl{tail}) comprises the fragments of all seven mentioned above phases, while it is evidently distinguished from a main body of the distribution.
\begin{figure*}
\centering
\includegraphics[width=0.47\textwidth]{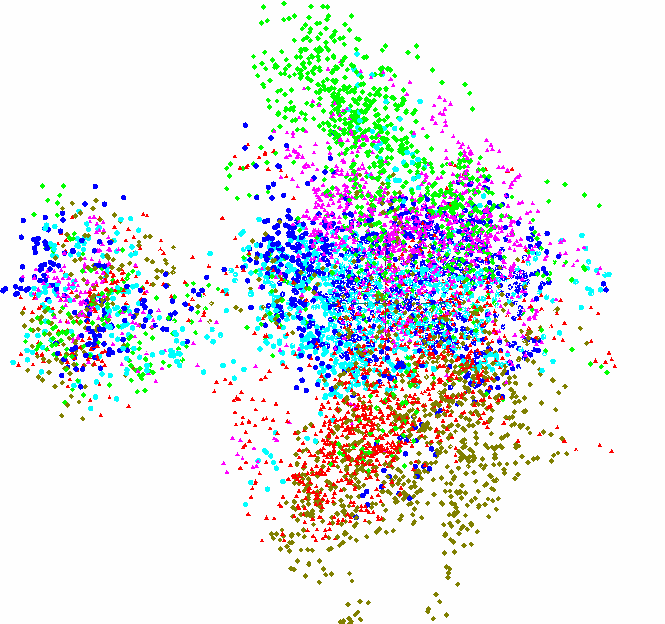}\hfill
\includegraphics[width=0.47\textwidth]{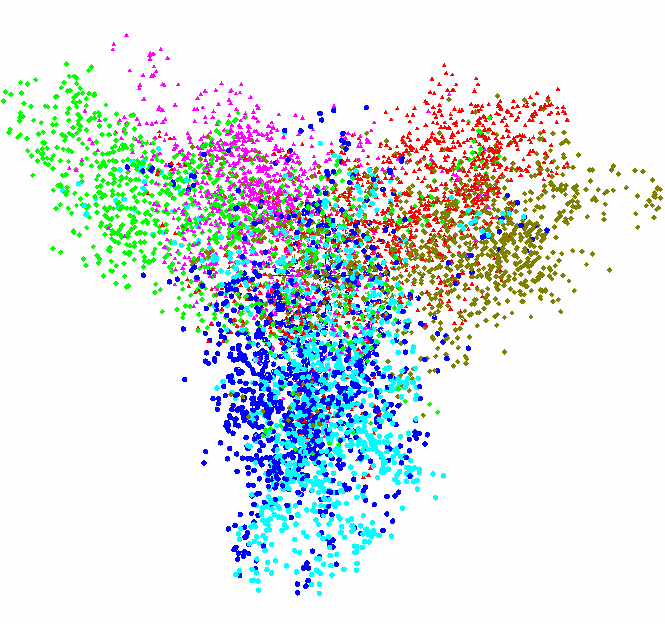}
\caption{{Same genome as in Fig.~\ref{fig1}, with \textsl{junk} phase erased; left is ``profile'' sight, and right is ``full face'' sight; color labeling is the same as in Fig.~\ref{fig1}.}}
\label{fig2}
\end{figure*}
This genome exhibits a typical ``four-cluster'' pattern, when seen in ``full face'': the clusters corresponding to leader and ladder strand coincide, in the this projection mode.

Another very important feature of this genome is the clearly visible \textsl{tail} phase, in the distribution of fragments. This is rather frequent pattern observed among the studied genomes. The difference between the dictionaries $W^{(0)}_{(3,3)}$, $W^{(1)}_{(3,3)}$ and $W^{(2)}_{(3,3)}$ (see Table~\ref{3slov}) manifests in the clustering in ``wings'' (shown in color in Figs.~\ref{fig1} and~\ref{fig2}); on the contrary, the lack of such difference observed for \textsl{junk} phase fragments results in a ball-shaped distribution of these points, in 63-dimensional space.
\begin{figure*}
\centering
\subfigure[\textbf{``profile''}]{\includegraphics[width=0.47\textwidth]{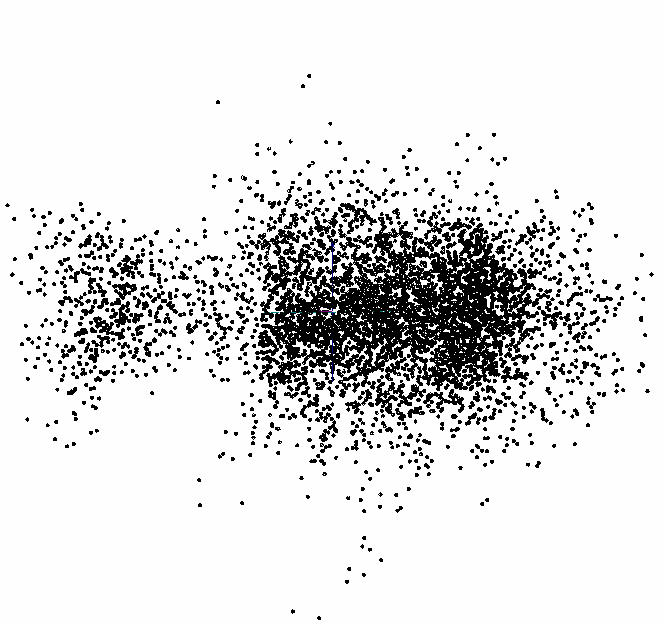}}\hfill
\subfigure[\textbf{``full face''}]{\includegraphics[width=0.47\textwidth]{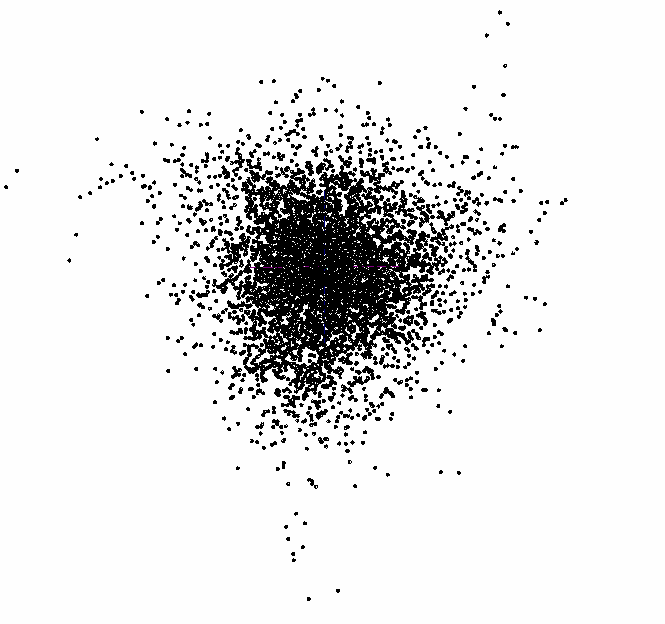}}
\caption{{The same genome as in Fig.~\ref{fig1}, \textsl{junk} phase shown in two projections, coding phases are erased.}}
\label{fig3}
\end{figure*}
Fig.~\ref{fig3} shows the \textsl{junk} phase fragments, only, with all the other erased from an image, in two projections. Unlike the bacterial genomes \cite{gorban1,gorban2}, here junk exhibits the separation into two clusters (see Fig.~\ref{fig3}(a)). It should be stressed that such two-cluster pattern of the distribution of the fragments falling into junk areas of a genome may not be observed, in general, without the fragments corresponding to coding areas of the genome. The question whether the \textsl{junk} phase fragments yield a pattern themselves, solely, is still open. Strictly speaking, this split of a junk fragments ensemble into a body of junk in proper sense, and in \textsl{tail} phase forces to claim the eight-cluster structure occurrence in chloroplasts genomes, in contrary to the patterns observed for bacterial genomes \cite{gorban1,gorban2}.

\begin{figure*}
\centering
\includegraphics[width=0.947\textwidth]{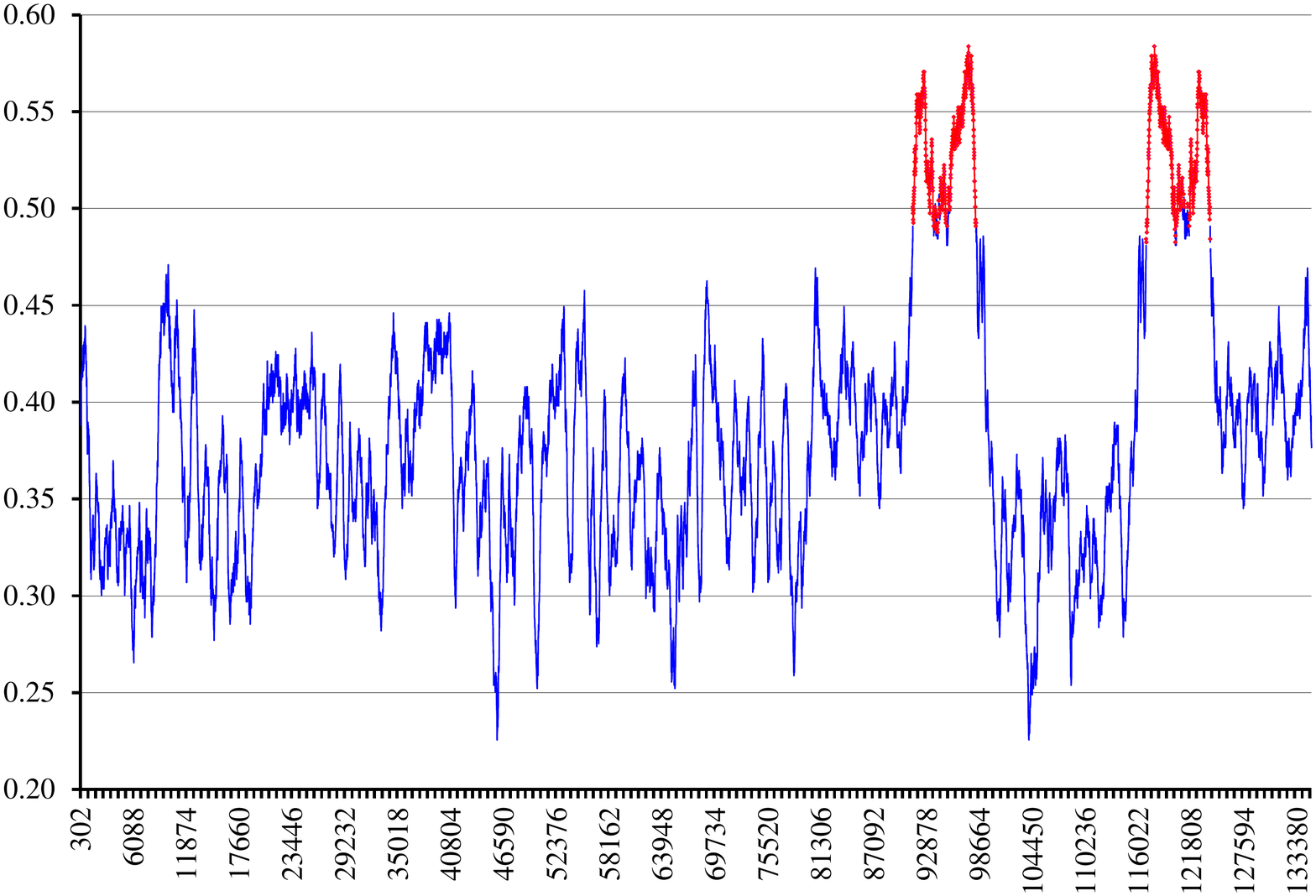}
\caption{{The plot of $\mathsf{GC}$-content of all the fragments layered out alongside the \textit{Lolium perenne} chloroplast genome; the \textsl{tail} phase (Fig.\,\ref{fig3}) is shown in red.}}
\label{fig4}
\end{figure*}

\begin{figure*}[t]
\centering
\includegraphics[width=0.95\textwidth]{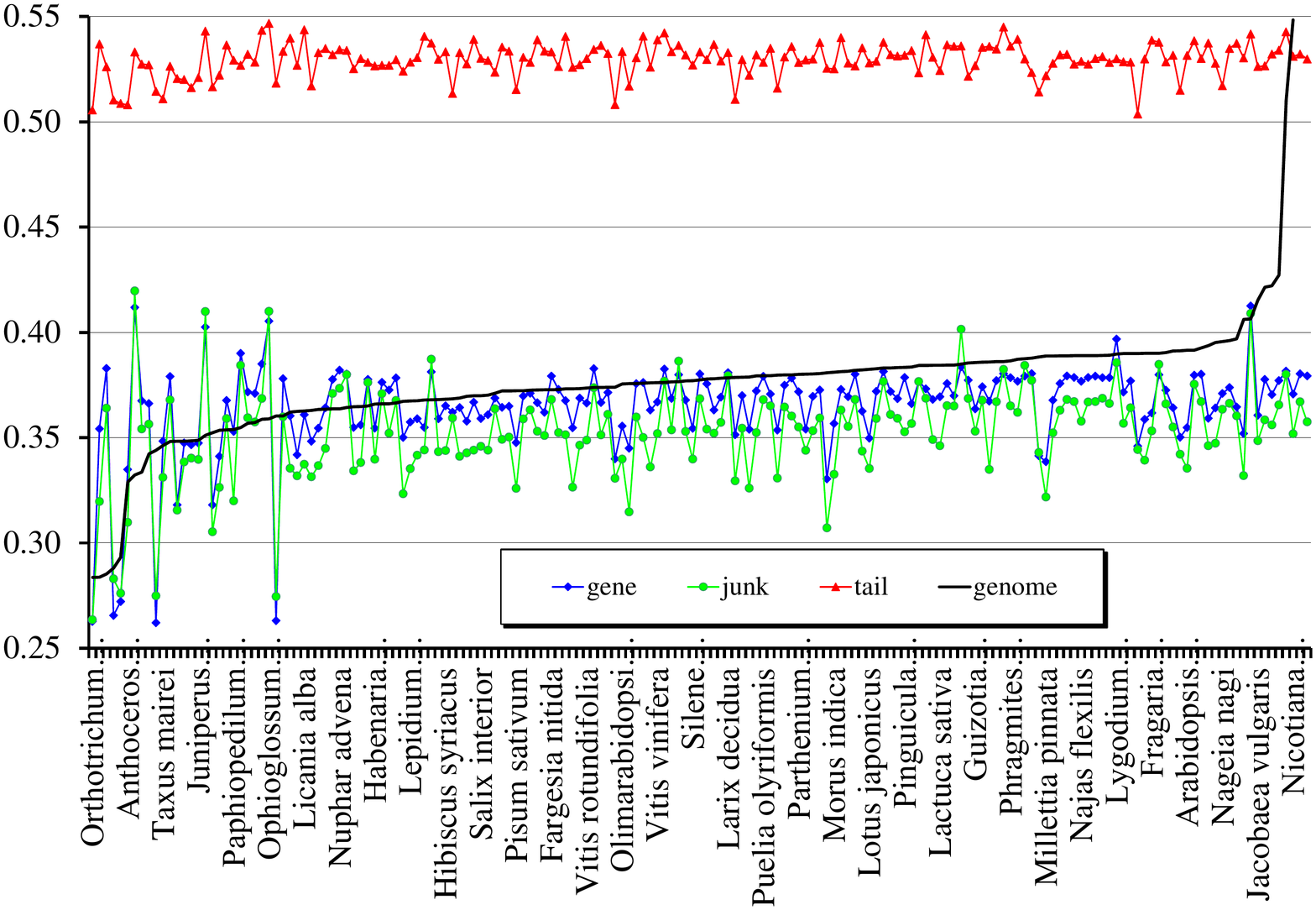}
\caption{{Average $\mathsf{GC}$-content for entire genome, coding, non-coding parts and \textsl{tail} phase (cumulatively).}}
\label{gc}
\end{figure*}

Let now consider the fragments comprising the tail in more detail. To do that, we determine $\mathsf{GC}$-content both for the entire genome, and for each fragment, and plot then the content against the number of a fragment. Fig.~\ref{fig4} shows this plot; the \textsl{tail} phase is colored in red. Let us remind, that the genome-wide $\mathsf{GC}$-content of this entity is equal to 0.38. The overall $\mathsf{GC}$-content has been reported to be the key factor defining the structure of clustering of the fragments formally identified within a bacterial genome \cite{gorban1,gorban2}; that former has significantly less effect, for chloroplast genomes. A tight examination of Table~\ref{genomy} shows that $\mathsf{GC}$-content varies from 0.28 (\textit{Orthotrichum rogeri}, AC~KP119739 and \textit{Syntrichia ruralis}, AC~FJ546412) to 0.51 for \textit{Selaginella moellendorffii}, AC~HM173080 and even 0,55 for \textit{Selaginella uncinata}, AC~AB197035. Meanwhile, this Table says nothing about the specific values of $\mathsf{GC}$-content of the fragments identified within various chloroplast genomes. Fig.~\ref{gc} answer this question.

This Figure shows the set of chloroplast genomes under consideration (see Table~\ref{genomy})
\begin{table}
\begin{tabular}{|l|c|c|c|}\hline
{}& Gene & junk & \textsl{tail}\\\hline
Genome & 0.9745 & 0.9617 & 0.6248\\\hline
Gene & & 0.9218 & 0,6285\\\hline
junk & & & 0.5939\\\hline
\end{tabular}
\caption{\label{tabcorr} Table of correlations}
\end{table}
ordered with respect to the genome-wide $\mathsf{GC}$-content value. Besides, this Figure shows the plots of average $\mathsf{GC}$-content determined over the ensemble of coding fragments (all six phases), of average $\mathsf{GC}$-content of non-coding fragments, and of average $\mathsf{GC}$-content of \textsl{tail} phase fragments. Evidently, the plots of genome-wide, coding and non-coding $\mathsf{GC}$-content figures exhibit a high concordance in behaviour, while the \textsl{tail} phase fragments ensemble remains rather permanent.

Table \ref{tabcorr} shows the correlations coefficients determined between averaged figures of $\mathsf{GC}$-content of four ensembles of the fragments of genomes. The figures shown in Table~\ref{tabcorr} reveal the relative independence of the \textsl{tail} phase from the other parts of a genome, and $\mathsf{GC}$-content of that latter never falls beyond 0.50 level. The set of genomes with lower figures of $\mathsf{GC}$-content comprises the species \textit{Orthotrichum rogeri}, \textit{Syntrichia ruralis}, \textit{Physcomitrella patens}, \textit{Marchantia polymorpha}, \textit{Sanionia uncinata}, \textit{Anthoceros angustus}, \textit{Ptilidiumpul cherrimum}, \textit{Equisetum arvense}, \textit{Glycyrrhiza glabra}, \textit{Trifolium subterraneum}, \textit{Orobanche gracilis}, \textit{Taxus mairei}, \textit{Millettia pinnata}, \textit{Pisum sativum}, \textit{Juniperus virginiana} and \textit{Juniperus bermudiana}. The genomes of \textit{Aneura mirabilis}, \textit{Lygodium japonicum}, \textit{Pteridium aquilinum}, \textit{Ophioglossum californicum}, \textit{Marsilea crenata} and \textit{Myriopteris lindheimeri} comprise the opposite group with higher figure of $\mathsf{GC}$-content. Finally, two species (these are \textit{Selaginella moellendorffii} and \textit{S. uncinata}) yield the highest level of $\mathsf{GC}$-content (see Table~\ref{genomy} for details).

Let now focus on the behaviour of the $\mathsf{GC}$-content of the fragments comprising \textsl{tails} in the distribution of the fragments (see Fig.~\ref{gc}). Remarkably, there is no genome with $\mathsf{GC}$-content figure lower than 0.5, for these fragments. Differing in this figure from the entire genome, the \textsl{tails} ensemble still comprises both coding, and non-coding fragments. The former are presented by densely located tRNA genes, 5S\,RNA and 16S\,RNA genes. This fact holds true for all genomes exhibiting a \textsl{tail}~phase, and such genomes make a majority of entities studied in this paper.

Let now provide some examples of the fragments distributions observed in chloroplast genomes with various $\mathsf{GC}$-content values. Consider the moss \textit{Physcomitrella patens} (AC\,AP005672) genome with $\mathsf{GC}$-content equal to 0.29 (next to the lowest one in the list of studied genomes).
\begin{figure*}
\centering
\includegraphics[width=0.47\textwidth]{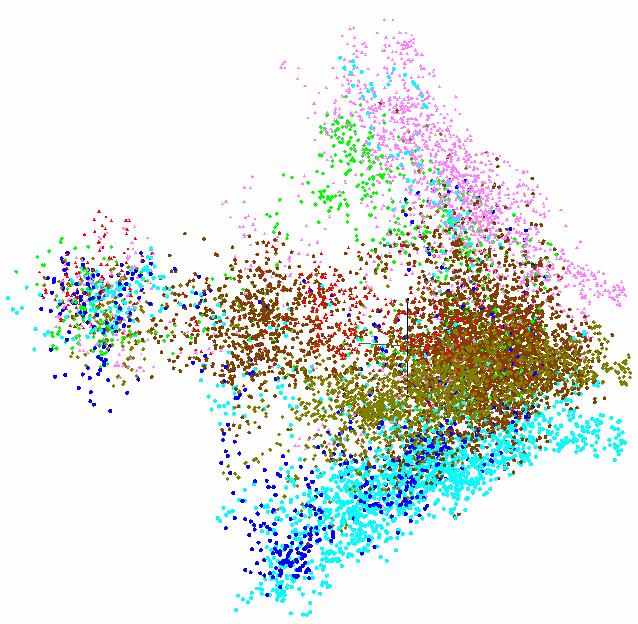}\hfill\includegraphics[width=0.47\textwidth]{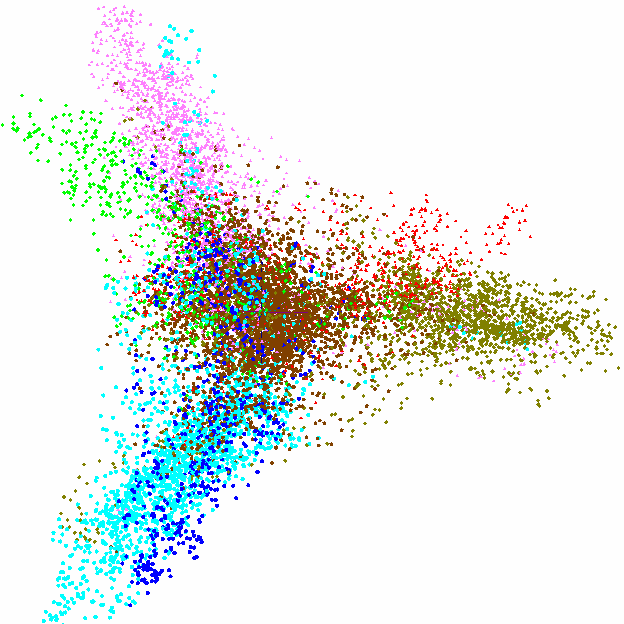}
\caption{{The distribution of 11\,118 fragments of moss \textit{Physcomitrella patens} chloroplast genome (AC\,AP005672), left is ``profile'', and right is ``full face'': $F_0$~-- red triangles, $F_1$~-- bright green diamonds, $F_2$~-- dark blue circles, $B_0$~-- rosy triangles, $B_1$~-- dark green diamonds, $B_2$~-- blue circles.}}
\label{fig_AP005672}
\end{figure*}
This is the model organism often used in evolutionary studies. Fig.~\ref{fig_AP005672} shows two projections of the full plot of the fragments; one easily can see
similar pattern with two ``tripods'' overlapping each other, and the \textsl{tail} phase part. It should be stressed, that this genome exhibits another triplet with the least standard deviation figure: $\mathsf{CGC}$, on the contrary to that one shown in Fig.~\ref{fig1}. This genome exhibits stronger split of two phases (these are $F_1$ vs. $B_0$ and $F_0$ vs. $B_1$), in comparison to the pattern shown in Fig.~\ref{fig2}; yet, the congruence of all the phases is strong enough.

To make it more clear, we show the distribution of all the fragments (the color labeling is the same as in Fig.~\ref{fig_AP005672}) falling in coding regions, only; all the points corresponding to \textsl{junk} phase are erased. This figure allows to see that the third phase in this genome slightly deviates, in its clustering pattern, from two other couples: a reasonable part of blue and dark blue points belong to another cluster than that one comprising the third phase fragments. This distribution is shown in Fig.~\ref{moss_no_junk}.

\begin{figure*}
\centering
\subfigure[\textbf{``profile''}]{\includegraphics[width=0.47\textwidth]{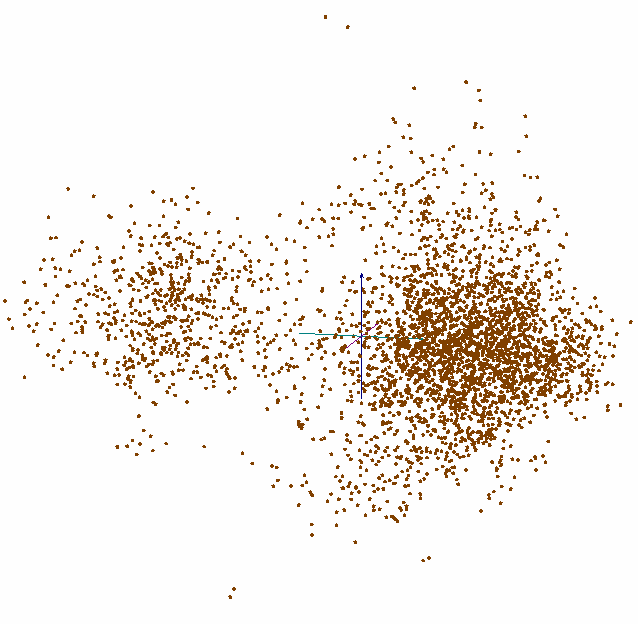}}\hfill
\subfigure[\textbf{``full face''}]{\includegraphics[width=0.47\textwidth]{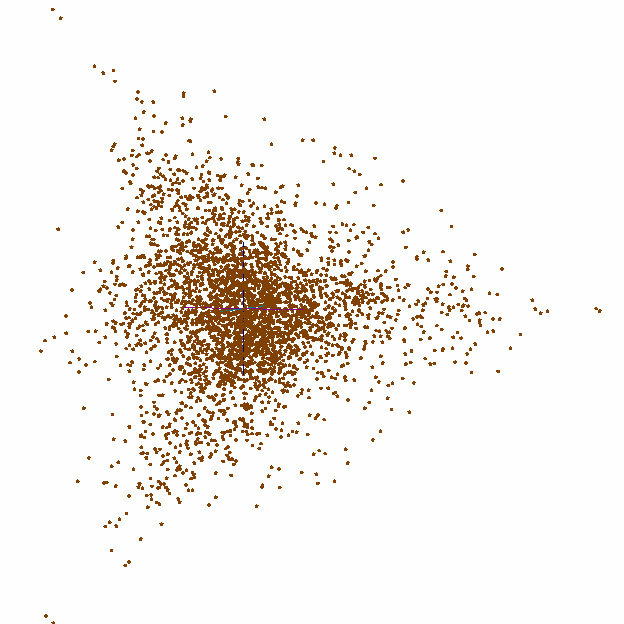}}
\caption{{\textsl{Junk}-phase only distribution of moss chloroplast genome fragments.}}
\label{moss_junk}
\end{figure*}

Fig.~\ref{moss_junk} shows the distribution of \textsl{junk} phase fragments of the moss genome. Similar to Fig.~\ref{fig3}, this genome also exhibits an occurrence of some points of \textsl{junk} in \textsl{tail} phase. Whether it is a biologically sounding fact, still awaits for an answer. Again, it should be borne in mind, that all the distributions shown in Figs.~\ref{fig_AP005672} to~\ref{moss_junk} are not independent: actually, all these figures just show the same distribution, while some points are not shown in various figures; still, they affect the distribution pattern.

\begin{figure*}
\centering
\subfigure[\textbf{``no junk, profile''}]{\includegraphics[width=0.45\textwidth]{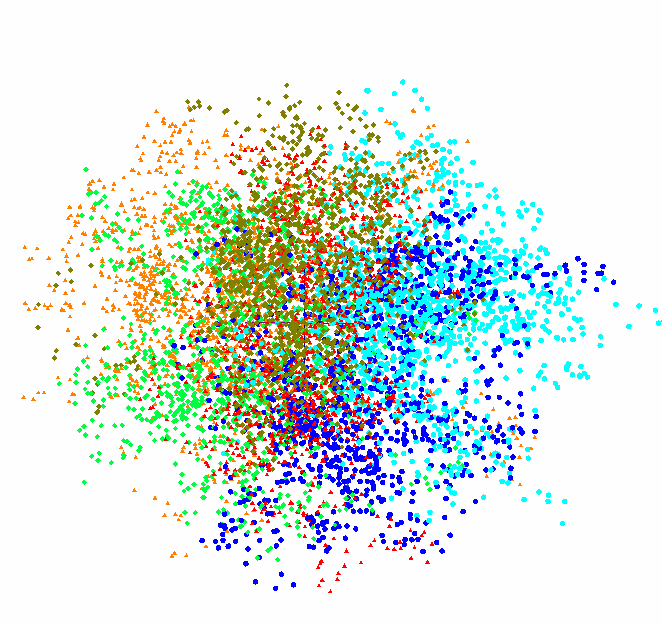}}\
\subfigure[\textbf{``junk, profile''}]{\includegraphics[width=0.45\textwidth]{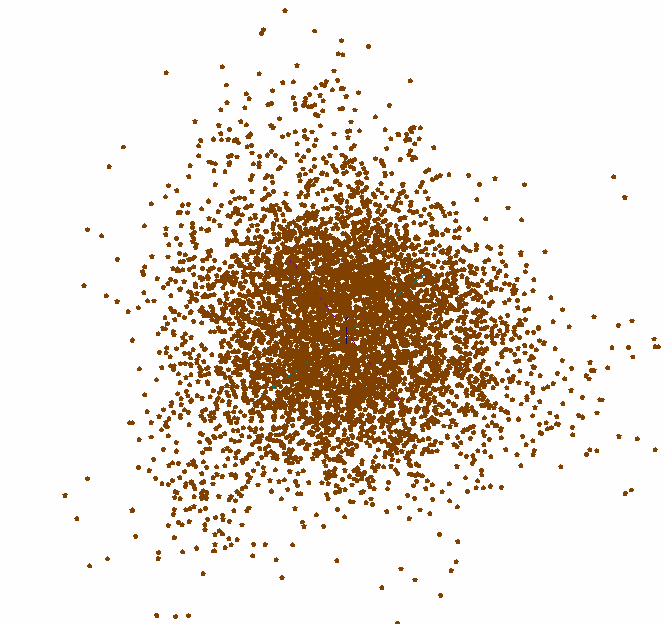}}\\
\subfigure[\textbf{``no junk, face''}]{\includegraphics[width=0.45\textwidth]{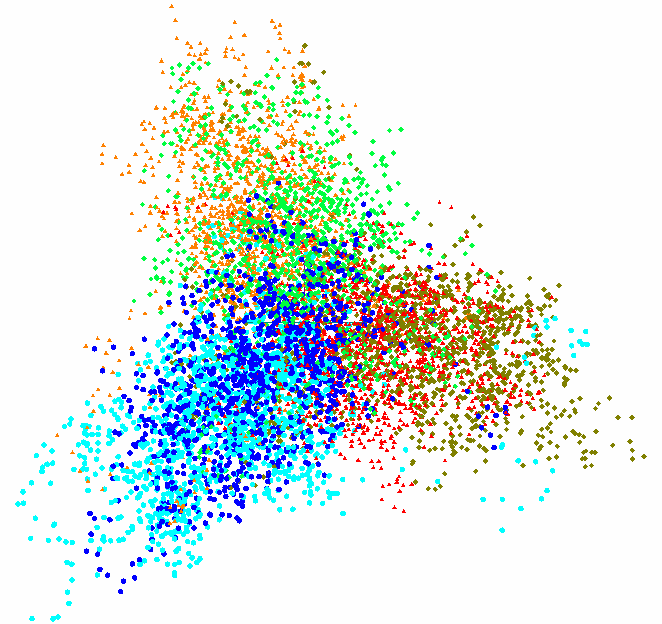}}\
\subfigure[\textbf{``junk, face''}]{\includegraphics[width=0.45\textwidth]{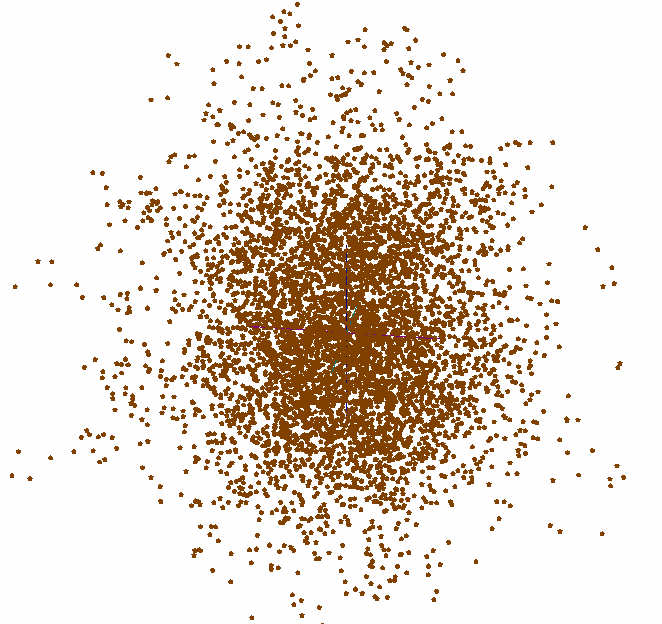}}
\caption{{\textit{Selaginella moellendorffii} chloroplast genome fragments distribution; upper is profile view, lower is face view.}}
\label{seginel1}
\end{figure*}

The patterns shown in Figs.~\ref{fig1} to~\ref{moss_junk} present a typical structuredness in a distribution of the small fragments of chloroplast genome. Actually, all the genomes except two entities exhibit such pattern in fragments distribution; these latter are the genomes of \textit{Selaginella moellendorffii} (AC\,HM173080) and \textit{S.\,uncinata} (AC\,AB197035). They are extremely ancient and rather isolated mosses belonging to primitive vascular plants. First of all, they have other triplets with the least standard deviation: $\mathsf{GAC}$ and $\mathsf{TAA}$, respectively. Fig.~\ref{seginel1} shows the distribution of all phases of \textit{S.\,moellendorffii} genome.
There is no \textsl{tail} phase at all, in this genome, neither in coding phases, nor in non-coding one. The pattern of distribution for \textit{S.\,uncinata} is pretty close to that one shown in Fig.~\ref{seginel1}. Another indirect evidence for this issue is discussed in \cite{cyano_ancest} (see also very useful paper \cite{sato}).

\subsection{Chloroplasts and cyanobacteria}\label{cianobak}
The essential difference in the structuredness of a genome of chloroplast from bacterial genome is the key issue of the work. Still, the question arises whether this difference is essential. In other words, while chloroplasts form a tight and uniform group of genome bearers, bacteria are extremely diverse, both in genetics, phylogeny, taxonomy, physiology and ecology. What if there are some bacteria that had fallen out from our analysis, but still are close to chloroplasts, in the sense of the small fragments distribution? Indeed, the diversity of bacteria is huge, and there is no guarantee of the total absence of the coincidence of the structure described above when retrieved from some bacterial genome.

\begin{figure*}
\centering
\subfigure[\textbf{``profile''}]{\includegraphics[width=0.47\textwidth]{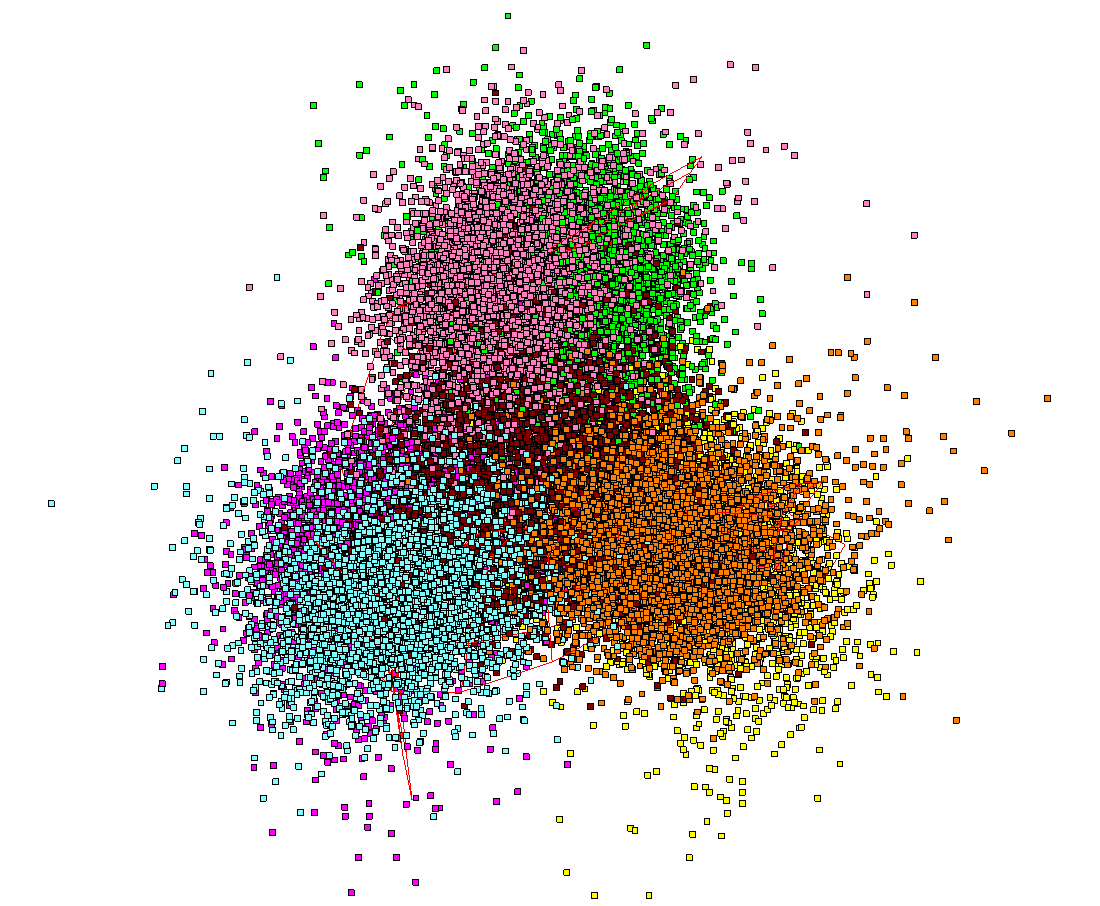}}\hfill
\subfigure[\textbf{``full face''}]{\includegraphics[width=0.47\textwidth]{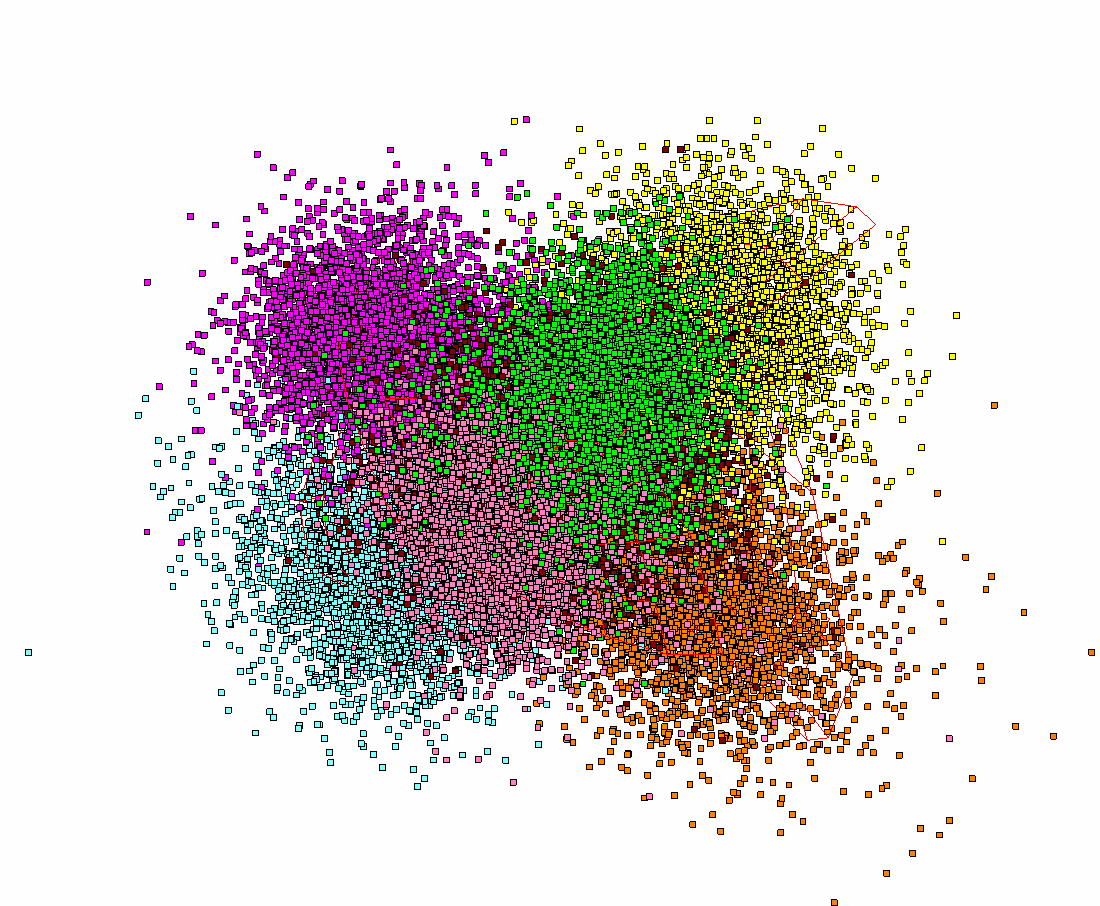}}
\caption{{\textit{Nostoc sp.} PCC 7107 distribution of fragments, $\Delta= 1005$, $R=202$.}}
\label{cyano}
\end{figure*}

Speaking on the similitude or any other semblance of the patterns observed in chloroplast genomes to those observed in bacterial genomes, one should first of all concentrate on the comparison of the structures of chloroplasts, and cyanobacteria. These latter are stipulated to be the other branch of descendants of the common ancestor of chloroplasts and modern bacteria. Here we do not study this point in detail, while some preliminary results \cite{zhob18} show that the divergence between chloroplasts and cyanobacteria is tremendous. Fig.~\ref{cyano} illustrate the point.

\section{Discussion}
Let now get back to the labeling system (see page~\pageref{junkdef}) of the formally identified fragments in a sequence. It provides a reasonable balance between the impact of coding and non-coding regions. Since the label value depends on the central nucleotide position, then approximately a half of the ``border'' fragments (i.\,e. those that cover the border between coding and non-coding regions in a genome) are labeled as \textsl{junk}, and another half are labeled as coding ones, with the specific phase value. Suppose, the total number of coding regions in a chloroplast genome is~50. Then an approximate number of ``border'' fragments labeled as \textsl{junk} is estimated as
\begin{equation}\label{bordernumber}
\dfrac{L}{2R}\times 50 \times 2 \approx 2\,500\,,
\end{equation}
where the factor 2 counts both forward and backward oriented coding regions. The same number~\eqref{bordernumber} of the ``border'' fragments would be labeled with some phase figures; this parity guarantees, to some extent, a lack of distortion in the fragments clustering.

Papers \cite{gorban1,gorban2} present an approach to figure out a structuredness in bacterial genomes based on systemic and sequential comparison of frequency dictionaries of the fragments of a genome; the fragments were identified in the same way, as we have done. It should be stressed that such fragments were identified with neither respect to a functional charge of a fragment. The results presented in these papers show that the fragments tend to cluster located in the vertices of two triangles. The triangle vertices correspond to the phase of a fragment; in other words, a cluster comprises the fragments that have the same reading frame shift figure. A mutual arrangement of these two triangles is completely determined by the average (over the genome) figure of $\mathsf{GC}$-content, for bacterial genomes.
\begin{figure}[h]
\centering
\includegraphics[width=0.47\textwidth]{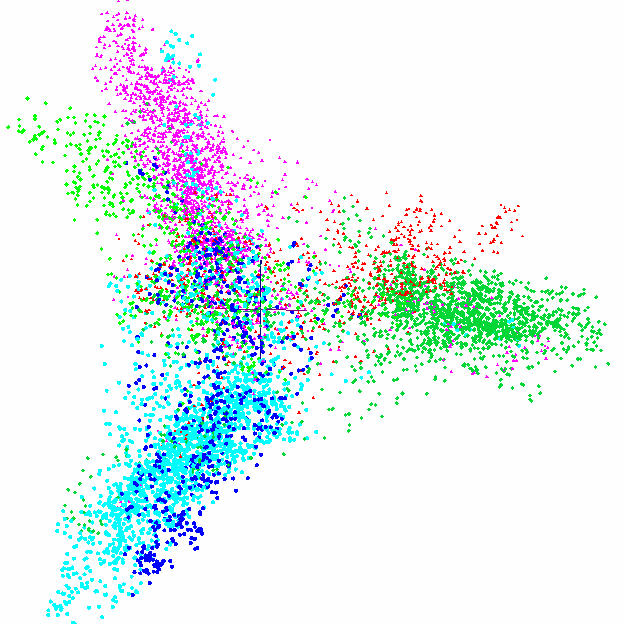}
\caption{\label{moss_no_junk} \textsl{Junk}-free distribution of moss chloroplast genome. Denotations are the same as in Fig.~\ref{fig_AP005672}.}
\end{figure}

A general seven-cluster structure was reported, for bacterial genomes, in these papers; the seventh cluster comprises the fragments falling into a junk area of the genome. The papers \cite{gorban1,gorban2} also provide an elegant explanation of an origin of this seven-cluster structuredness, describing the constraints forcing two triangles to rotate and project one over another. Here the genome-wide $\mathsf{GC}$-content is claimed to be the only key factor determining the pattern of the cluster structure. A minor variation of $\mathsf{GC}$-content results in visible change of the structure pattern.

There are following patterns of the fragments distribution, observed on bacterial genomes, for various figures of $\mathsf{GC}$-content. $\mathsf{GC}$-content close to 25\,\% yields two ``parallel triangles'' (for $\mathsf{AT}$-reach genomes); the growth of $\mathsf{GC}$-content to $\sim 35\,\%$ yields the pattern with two ``orthogonal triangles'', and the raise of $\mathsf{GC}$-content up to 60\,\% results in degeneration of two triangles into a single plane. Besides, the authors of \cite{gorban1,gorban2} claim such seven-cluster pattern be universal one; meanwhile, our results disprove this hypothesis.

\subsection{Cluster structure of chloroplast genomes}
Since chloroplasts take their origin from bacteria \cite{merezh,merezh1,curopin2014,ravenallen}, then one may expect they inherit this universal pattern of the inner genome structuredness. Our results disprove this assumption; moreover, $\mathsf{GC}$-content of chloroplast genomes does not impact on the pattern of fragments distribution. The newly found pattern in small fragments distribution in 63-dimensional triplet frequency space seems to be very universal: there are two only exclusions from the list of studied genomes (see Table~\ref{genomy}). They are presented by two ancient moss species (\textit{Selaginella moellendorffii} and \textit{Selaginella uncinata}) originated more than $4\times 10^{8}$ years ago.

The list of other genomes (see Table~\ref{genomy}) is apparently split into two parts: the former has the triplet $\mathsf{GCG}$ exhibiting the least standard deviation level, and the latter has the triplet~$\mathsf{CGC}$. Indeed, these triplets exhibit a high similitude. The point is that they comprise so called \textsl{complementary palindrome}. Palindrome itself is a word that read equally in opposite directions (e.\,g., \textsl{level} in English; two words may form a palindromic couple (\textsl{god} $\Leftrightarrow$ \textsl{dog} in English). The triplets~$\mathsf{GCG}$ and~$\mathsf{CGC}$ yield the so called complementary palindrome: a couple of two strings of DNA sequence that are read equally in opposite directions with respect to the Chargaff's substitute rule ($\mathsf{A} \Leftrightarrow \mathsf{T}$ and $\mathsf{C} \Leftrightarrow \mathsf{G}$). This really important symmetry in frequency dictionaries, but it completely falls beyond the scope of this paper.

Another important question here is whether the observed clusters corresponding to six phases (these are $F_0$, $F_1$, $F_2$, $B_0$, $B_1$ and $B_2$) really comprise clusters, or it is a kind of artifact resulted from a visualization technique. This question has obvious and transparent answer: yes, the clusters observed by visualization of the phases are the real clusters identified with a clustering technique. To check it, we have carried out $K$-means cluster implementation, of the frequency dictionaries corresponding to the fragments.
\begin{figure*}
\centering
\subfigure[\textbf{``full face''}]{\includegraphics[width=0.47\textwidth]{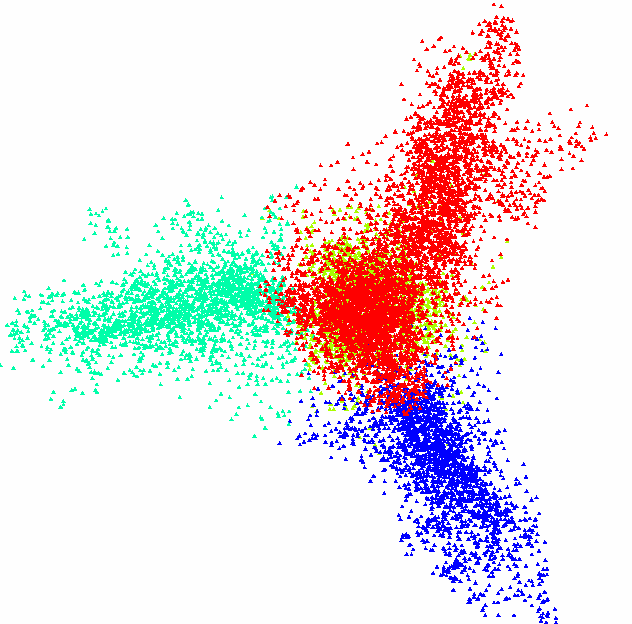}}\hfill
\subfigure[\textbf{inner coordinates}]{\includegraphics[width=0.47\textwidth]{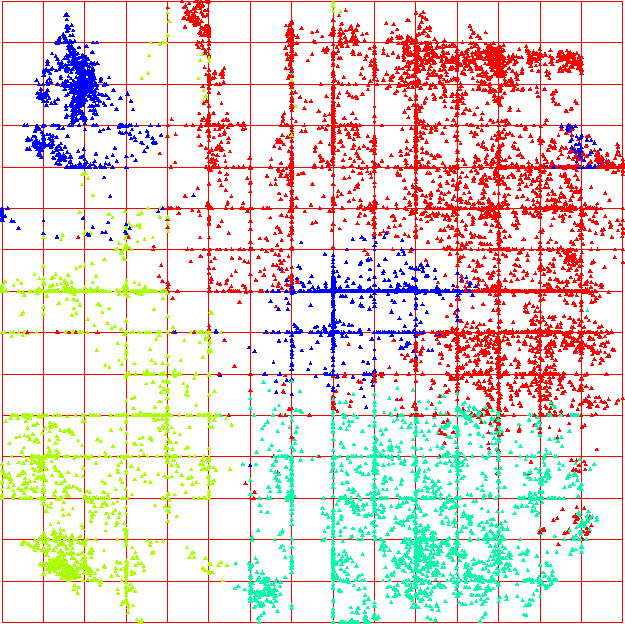}}
\caption{{$K$-means ($K=4$) for \textit{Physcomitrella patens} (AC\,AP005672) chloroplast genome.}}
\label{4klassa}
\end{figure*}
Fig.~\ref{4klassa} shows the clustering developed by $K$-means \cite{fukunaga} (with $K=4$) for the moss genome. Again, we did not aim to figure out some cluster structure due to $K$-means, but to verify the cluster structure observed in genomes through the visualization (that is the phase coloring). The clustering shown in Fig.~\ref{4klassa} is very stable: a hundred runs of $K$-means always resulted in the same distribution of points. So, the clusters like those shown in Figs.~\ref{fig1}, \ref{fig2}, \ref{fig_AP005672} and~\ref{moss_no_junk} are the really existing entities, not an artifact.

Few words should be said towards the pictures shown in Fig.~\ref{4klassa}. The left picture presents the distribution of all the fragments (of course, converted into frequency dictionaries) in 63-dimensional space, in principal components, the ``full face'' projection and clustered into four classes by $K$-means. Obviously, the classes identified by $K$-means comprise the points belonging both to some coding phase, and to the non-coding phase; yet, we did not aim to separate the points by $K$-means in the same manner, as by coloring. The right picture shows the same distribution in so called inner coordinates of an elastic map; the details on this techniques could be found in \cite{n3_3,n4,gorbanzyn2,gorbanzyn,gorbanzyn1,gorbanzyn3}.

Careful examination of Figs.~\ref{fig1}, \ref{fig2}, \ref{fig_AP005672} and~\ref{moss_no_junk} shows the general situation in localization of the phases, within a pattern. Indeed, the localization of the phases could be described by the following cyclic diagrams: $F_0 \rightarrow F_1 \rightarrow F_2 \rightarrow F_0$ (clockwise), and $B_0 \rightarrow B_1 \rightarrow B_2 \rightarrow B_0$ (counterclockwise). In fact, these two diagrams mirror each other, so that no complete coincidence \begin{figure*}
\centering
\subfigure[\textbf{``up''}]{\includegraphics[width=0.47\textwidth]{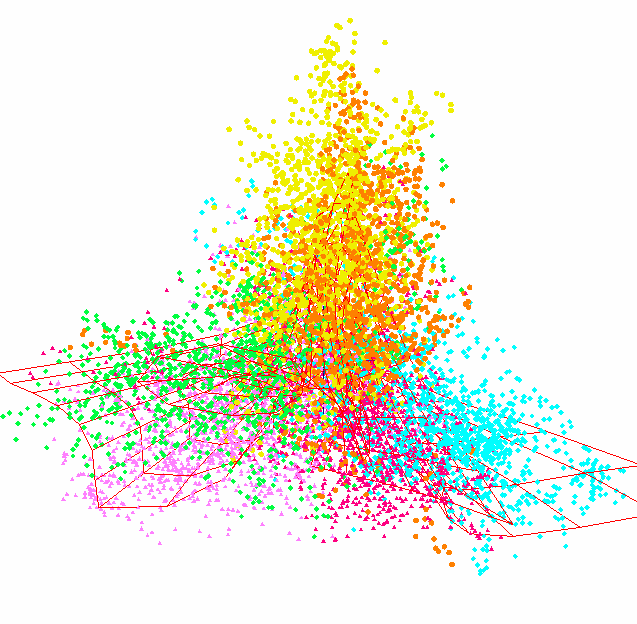}}\hfill
\subfigure[\textbf{``down''}]{\includegraphics[width=0.47\textwidth]{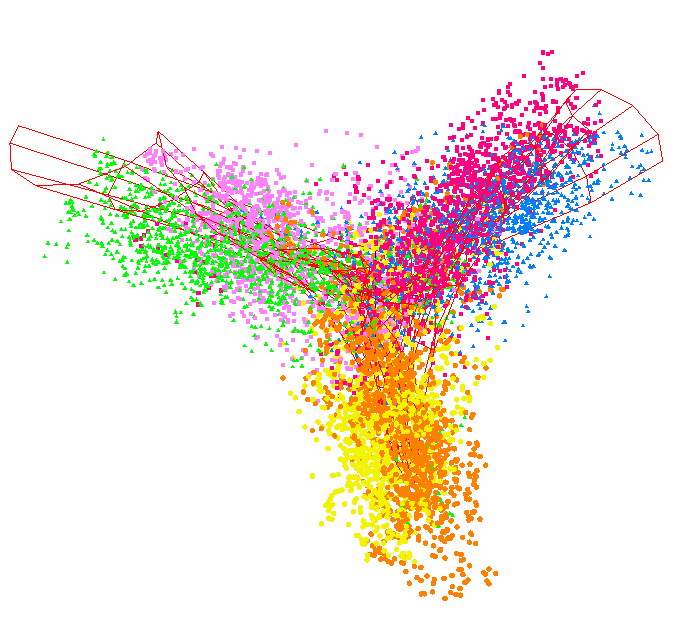}}
\caption{{``Up'' and ``down'' orientation of the clusters shown on two genomes: \textit{Ranunculus macranthus} (up), and \textit{Anthoceros angustus} (down).}}
\label{upanddown}
\end{figure*}
make take place due to rotation. Such mirror symmetry corresponds to the double-stranded structure of DNA; the localization of~$F_2$ and~$B_2$ phases in the same projection is here of greater importance. All the studied chloroplast genomes exhibit such mirroring symmetry, while there are no evidences for that latter in bacterial ones \cite{gorban1,gorban2}. The phases $F_0$, $F_1$, $F_2$ make a triangle with given vertices circuit direction; same is true for the phases $B_0$, $B_1$ and $B_2$, and the circuit direction is the same, as for $F$-phases. This fact seems to be universal for bacteria (and some other genomes, e.\,g. fungi ones); on the contrary, chloroplast genomes exhibit exactly opposite pattern: they have counter-directed circuit directions, for those phases. Papers \cite{assym_pnas,struktura} report on another type of structuredness found in chloroplast genomes, while we believe the mechanism staying behind these structures and those we are showing here, is the same: triplet frequency peculiarities. More specific mechanism based on codon bias yields a structuredness reported in \cite{triplsymmetry}. These facts may reveal the ``independent'' evolution of chloroplast genomes (see also \cite{comparanal,chloro_evol2}), on the contrary to the synchronized evolution of these latter with the host nuclear genome \cite{nasha2}. Also, such symmetry may answer the question towards the attribution of contigs for \textsl{de novo} assembling genomes \cite{sfu1,sfu2,sfu3} (see also another sight on the problem in \cite{necodir_evol}).

This mirroring has one more manifest in mutual location of the clusters comprising different phases. Fig.~\ref{upanddown} illustrates this fact: while the location of phase~0 and phase~1 remains the same, in both subfigures, the location of the phase~2 takes mirroring positions. The phase~2 cluster faces down, for \textit{Anthoceros angustus}, and that former faces up for large buttercup (\textit{Ranunculus macranthus}). To make the images more apparent, we have erased the points corresponding to junk. Two positions of phase~2 cluster correspond to two mirroring axes systems. Let now get back to Table~\ref{genomy}; the last column in the table (labeled with $\Updownarrow$ sign) indicates the orientation of the clusters: $\mathsf{U}$ stands for ``up'' positioning of the cluster, and $\mathsf{D}$ stands for ``down'' positioning of that latter. Comparing Figs.~\ref{fig2}, \ref{moss_no_junk}, \ref{fig_AP005672}, \ref{upanddown} to Fig.~\ref{cyano} (see Subsec.~\ref{cianobak}), one sees that such mirroring symmetry is universal, for chloroplast genomes; cyanobacteria that are claimed to be evolutionary related to chloroplasts, do not exhibit such pattern, at all.

Another sounding manifestation of the symmetry is the interchange of the triplet yielding the least standard deviation figure; see again Table~\ref{genomy}. Indeed, with exclusion of two triplets (these are $\mathsf{GAC}$ and $\mathsf{TAA}$), all other entries exhibit either $\mathsf{GCG}$, or $\mathsf{CGC}$ triplet yielding the least standard deviation figure. The species with unconventional triplets $\mathsf{GAC}$ and $\mathsf{TAA}$ are actually the ancient moss organisms of \textit{Selaginella} genus appeared app.~400 billion years ago. The unconventionality of the triplets yielding the least standard deviation figure may result from this long isolated lineage.

\begin{table}
\begin{tabular}{|c|c|c|}\hline
{} & $\mathsf{CGC}$& $\mathsf{GCG}$\\\hline
$\mathsf{U}$ & 41 & 95 \\\hline
$\mathsf{D}$ & 20 & 19 \\\hline
\end{tabular}
\caption{\label{U+D} Distribution of patterns.}
\end{table}
Apart these two species, all other ones (see Table~\ref{genomy}) could be split into two groups: the former with $\mathsf{GCG}$ triplet yielding the least standard deviation, and the latter with $\mathsf{CGC}$ triplet; the abundances of each groups are 115 and 61 entries, respectively. It should be mentioned that two genomes were not annotated, completely; thus, we were not able to determine what type of symmetry they exhibit. Table~\ref{U+D} summarizes the distribution of chloroplast genomes over the combinations of $\mathsf{U} \leftrightarrow \mathsf{D}$ variants, and the triplets $\mathsf{CGC} \leftrightarrow \mathsf{GCG}$. In such capacity, the genomes with $\mathsf{CGC}$ triplet differ from those with $\mathsf{GCG}$ ones. Whether this difference is of a nature of things, or results from a bias of the database used in the study, should be examined further. One definitely could say there is no correlation between the pattern of orientation, triplet with the least standard deviation figure, and separation of plants on gymnosperm vs. angiosperm species (cf. Table~\ref{U+D} and Table~\ref{genomy}).

\subsection{Specific type of symmetry and coding regions}
Consider now the abundances of the beams (or clusters) corresponding to the phases~$F_0$ through~$B_2$; obviously, the must be equal, or at least pretty close, since the beams differ in the reading frame shift of a triplet, only. Typical figures are the following: $|F_0+B_0| = 2489,4$, $|F_1+B_1| = 2488,6$, and $|F_2+B_2| = 2485,8$. The greatest standard deviation of the beam abundances is provided by \textit{Hibiscus syriacus} (AC\,KP688069 in EMBL--bank), and the figure is~14.53. Reciprocally, the least figure (that is exactly zero) is provided by \textit{Olimarabidopsis pumila} (AC\,AP009368).

\begin{figure*}
\centering
\includegraphics[width=1.01\textwidth]{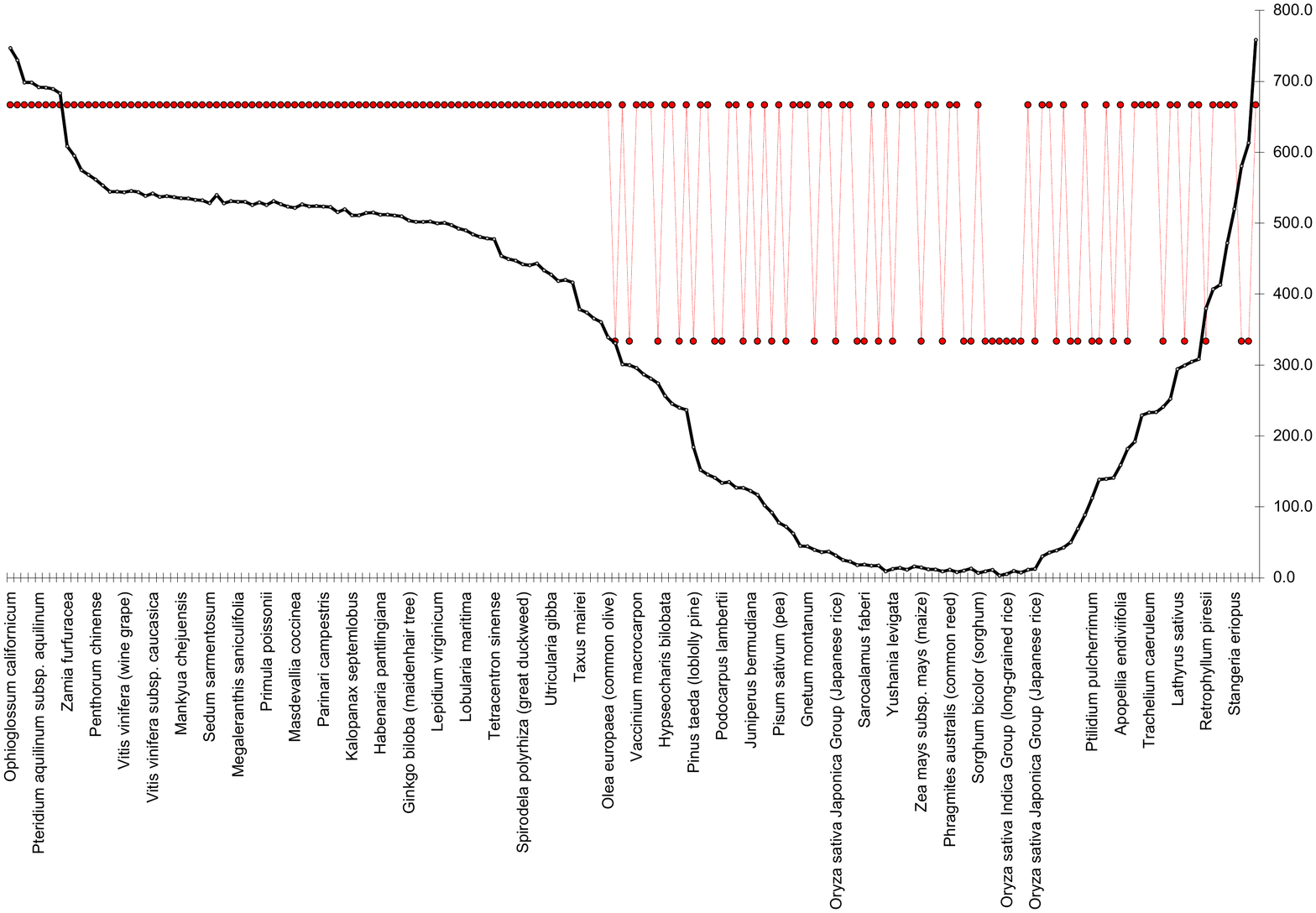}
\caption{{The bias within a phase, and the type of mirroring symmetry; see explanation in text.}}
\label{dif}
\end{figure*}

The difference between the phases $|F_0-B_0|$, $|F_1-B_1|$ and $|F_2-B_2|$ are of greater interest. These figures vary from $-1305$ (averaged over three beams), for \textit{Ophioglossum californicum} (AC\,KC117178) to 1387, for \textit{Equisetum arvense} (AC\,GU191334). Fig.~\ref{dif} shows the relation of the bias in forward and backward coding regions occurrence, in different organisms, and the type of their mirroring symmetry.
This figure shows the set of genomes (Table~\ref{genomy}) ordered ascending on $|F_0-B_0|$ figures; in other words, the left genome has $|F_0| - |B_0| = -1362$ (that is \textit{Ophioglossum californicum}, AC\,KC117178), while the right one exhibits $|F_0| - |B_0| = 1382$ (that is \textit{Equisetum arvense}, AC\,GU191334). The solid black line in this Fig. shows the standard deviation of cluster abundances determined over all six phases; small red diamonds show the symmetry orientation: upper dots show $\mathsf{U}$ type, and lower ones show $\mathsf{D}$ type. It seems that the excess of the abundance of the fragments belonging to backward phases over those belonging to forward phases in 600 entities results in the unambiguous determination of $\mathsf{U}$ type symmetry orientation; the right end of this curve supports, to some extent, this idea.

\subsection{What is beyond?}
The study of statistical properties of DNA sequences still challenges researchers, and may bring a lot new. Here we have demonstrated basic structural difference of chloroplast genomes from the bacterial ones, manifested in the clustering in distribution of small formally identified fragments of a genome. Below are some issues that had fallen from the scope of this paper, while they are rather important to be considered in the nearest future.

\subsubsection{Dark matter of a genome}
Functional and evolution roles of the junk in a genome still is conspired from researchers. It is extremely doubtful that junk has no matter in a genome, neither it plays no role in various and complicated biological processes involved into an inherited information processing and functioning. For some cases (see \cite{sfu1,sfu2,sfu3}, the removal of junk enforces the clustering of coding regions and makes easier the comprehension of the peculiarities standing behind. Yet, special efforts must be addressed to reveal the role and impact of junk regions of a genome on the processes mentioned above.

A variety of aspect of the influence of a junk on clustering observed within a genome is very wide. Not speaking about the differences in statistical properties of frequency dictionaries $W_{(3,3)}$ (and $W_{(m,n)}$, in general) observed for junk fragments of a genome vs. those observed for coding ones, one may expect even the strong impact from the ratio of coding/non-coding parts occurred within a genome. For instance, here we report on mirror symmetry in mutual interlocation of six coding phases, for the frequency dictionary~$W_{(3,3)}$ developed for chloroplast genomes. Fig.~\ref{cyano} explicitly demonstrates an absence of such symmetry, for cyanobacteria genome, and this fact may result from a significant difference in the coding/non-coding ratio figures observed for these genetic systems.

\subsubsection{Other chloroplast genomes}
Here we present some results obtained on the careful examination of 178 genomes of ground plants. Yet, the generality of the observation awaits for further approval: first of all, one should study the chloroplast genomes of the organisms deviating rather far from the studied ones, in ecology (water plants, and algae, especially), physiology, taxonomy. Such systemic examination is the matter of the nearest future work of ours.

\section{Conclusion}
Here we studied the structuredness of chloroplast genomes revealed trough the clustering of frequency dictionaries of considerably short fragments of a genome that were determined formally, with neither respect to the function encoded in a part of the genome fell into the fragment. The triplet dictionaries were developed, to cluster; these former counts triplets with no overlapping, while with no gaps between any two triplets. The fragments are distributed into eight distinct clusters: six of them gather the fragments falling into the coding regions, and differ in reading frame shift; the shift manifests in phase index of a fragment. The seventh cluster comprises the fragments falling into non-coding regions, and finally, he eighth cluster (so called \textsl{tail}) comprises the fragments with excessive $\mathsf{GC}$-content value. These fragments correspond to the region where various tRNA and S\,RNA genes are concentrated; probably, this cluster includes also the ``border'' fragments (those that contain a border between coding and non-coding parts of a genome).

The clusters exhibit wonderful mirroring symmetry: the phase circuit in the forward and backward strands are counter-directed; this fact completely contradicts to the similar structure observed for bacteria, including cyanobacteria (which are stipulated to be the descendants of a common ancestor with chloroplasts). Such mirror symmetry yields a separation of the genomes into two groups: those with ``up''-directed location of the cluster comprising $F_2$ and $B_2$ phases vs. those with ``down''-directed; apparently, the threshold in the abundances of the phases gathered into a single cluster determines the direction of the $F_2 \div B_2$ cluster.

\begin{acknowledgments}
We would like to extend our gratitude to Prof.~Alexander~N.\,Gorban from Leicester University and Andrew Yu.\,Zinovyev from Curie Institute for long-time collaboration and permanent encouraging interest to the work. The work was partly supported by the grant from Russian Government (\#~14.Y26.31.0004).
\end{acknowledgments}

\end{document}